\documentclass{llncs}

\usepackage{times}

\usepackage{llncsdoc}
\usepackage{amsmath}
\usepackage{amssymb}
\usepackage{latexsym}
\usepackage{algorithm}
\usepackage{algorithmic}
\usepackage[dvips]{graphicx,color}
\usepackage{setspace}

\usepackage{amsfonts}
\usepackage{enumerate}
\newtheorem{defn}{Definition}

\newcommand{\ti}[1]{\textit{#1}}
\newcommand{\tbf}[1]{\textbf{#1}}
\newcommand{\mc}[1]{\mathcal{#1}}

\newcommand{\Blacksquare}{\rule{.7em}{.7em}}
\newcommand{\epf}{\hfill $\Blacksquare$}

\newcommand{\lt}{{{\L}o{\'s}-Tarski}}

\newcommand{\PS}{\ensuremath{\mathbb{PS}}}
\newcommand{\PSC}{\ensuremath{\mathbb{PSC}}}
\newcommand{\PSCf}{\ensuremath{\mathbb{PSC}_f}}

\newcommand{\PE}{\ensuremath{\mathbb{PE}}}
\newcommand{\PCE}{\ensuremath{\mathbb{PCE}}}
\newcommand{\PCEf}{\ensuremath{\mathbb{PCE}_f}}

\begin{document}

\title{Generalizations of the {\lt}\\ Preservation Theorem}

\author{Abhisekh Sankaran, Bharat Adsul, Supratik Chakraborty}
\institute{Indian Institute of Technology (IIT), Bombay, India\\
\{abhisekh, adsul, supratik\}@cse.iitb.ac.in}

\maketitle
\thispagestyle{empty}

\begin{abstract}
\noindent 
We present new preservation theorems that semantically characterize
the $\exists^k \forall^*$ and $\forall^k \exists^*$ prefix classes of
first order logic, for each natural number $k$.  Unlike preservation
theorems in the literature that characterize the $\exists^* \forall^*$
and $\forall^* \exists^*$ prefix classes, our theorems relate the
count of quantifiers in the leading block of the quantifier prefix to
natural quantitative properties of the models.  As special cases of
our results, we obtain the classical {\lt} preservation theorem for
sentences in both its extensional and substructural versions. For
arbitrary finite vocabularies, we also generalize the extensional
version of the {\lt} preservation theorem for theories. We also
present an interpolant-based approach towards these results. Finally,
we present partial results towards generalizing to theories, the
substructural version of the {\lt} theorem and in the process, we give
a preservation theorem that provides a semantic characterization of
$\Sigma^0_n$ theories for each natural number $n$.
\end{abstract}


\section{Introduction}

Preservation theorems in first order logic (henceforth called FO) have
been extensively studied in model theory. A FO preservation theorem
for a model-theoretic operation syntactically characterizes FO
definable classes of structures that are preserved under that
operation.  A classical preservation theorem (also one of the
earliest) is the {\lt} theorem, which states that over arbitrary
structures, a FO sentence is preserved under substructures iff it is
equivalent to a universal sentence~\cite{chang-keisler}.  In dual
form, the theorem states that a FO sentence is preserved under
extensions iff it is equivalent to an existential sentence.  It is
well-known that if the vocabulary is relational, the sizes of the
minimal models of a sentence preserved under extensions are no larger
than the number of quantifiers in an equivalent existential sentence.
Thus, the dual version of the {\lt} theorem not only asserts the
equivalence of a syntactic and a semantic class of FO sentences, but
also yields a relation between a quantitative model-theoretic property
(i.e., sizes of minimal models) of a sentence in the semantic class
and the count of quantifiers in an equivalent sentence in the
syntactic class.

Counts of quantifiers are known to have a bearing on the parameterized
complexity, and even decidability, of satisfiability checking of
various syntactic classes of FO~\cite{graedel}.  For example, consider
the class of prenex FO sentences, over a relational vocabulary, having
a prefix structure of the form $\exists^*\forall^*$.  It is known that
satisfiability checking for this class is in
$\mathsf{NTIME}((nk^m)^c)$, where $n$ is the length of the sentence,
$k$ and $m$ are the number of existential and universal quantifiers
respectively in the sentence, and $c$ is a suitable
constant~\cite{graedel}.  Similarly, for each $k \ge 2$,
satisfiability checking for the class of sentences of the form
$\forall^k \exists^* \varphi$ where $\varphi$ is quantifier-free, is
undecidable if equality is allowed in $\varphi$~\cite{graedel}.  It is
therefore interesting to study preservation theorems that relate
quantitative properties of models of sentences in a semantic class to
counts of quantifiers of equivalent sentences in a syntactic class.
In recent years, there has been significant interest in syntactic
classes of FO with one quantifier alternation in the context of
program verification, program synthesis and other
applications~\cite{srivastava,emmer,gulwani}.  The literature contains
several semantic characterizations for these syntactic classes using
notions such as ascending chains, descending chains, and
1-sandwiches~\cite{chang-keisler} (also see Appendix
\ref{appendix:comparison}). However, none of these relate quantifier
counts to any model-theoretic properties. In this paper, we take a
step towards addressing this problem.  Specifically, we present
preservation theorems that provide new semantic characterizations of
sentences in prenex normal form with quantifier prefixes of the form
$\exists^k\forall^*$ and $\forall^k\exists^*$. Our theorems relate the
count $k$ of quantifiers \emph{in the leading block} of quantifiers to
quantitative properties of the models.

The present work builds on notions introduced earlier
in~\cite{wollic12-paper}, specifically, those of \emph{cores} and
\emph{substructures modulo bounded cores}. It was conjectured
in~\cite{wollic12-paper} that for every natural number $k$, a FO
sentence is preserved under substructures modulo $k$-sized cores iff
it is equivalent to a prenex sentence with quantifier prefix of the
form $\exists^k\forall^*$.  In this paper, we formally prove this
conjecture over arbitrary structures.  This gives us a preservation
theorem that generalizes the substructural version of the {\lt}
theorem for FO sentences.  Our proof approach consists of introducing
a notion dual to that of substructures modulo $k$-sized cores, and
then proving a dual version of the original conjecture.
Interestingly, the dual version of the conjecture leads to a
generalization of the extensional form of the {\lt} theorem for
sentences. To the best of our knowledge, our characterizations are the
first to relate natural quantitative properties of models of sentences
in a semantic class to the count of quantifiers in equivalent
$\exists^* \forall^*$ or $\forall^* \exists^*$ sentences.  For
arbitrary finite vocabularies, we also generalize the extensional
version of the {\lt} theorem for theories.  We also present
interpolant-based semantic characterizations of the same syntactic
classes as considered above. Finally we present our partial results
towards generalizing the substructural version of the {\lt} theorem
for theories over arbitrary finite vocabularies.

The rest of the paper is organized as follows. In Section
\ref{section:PSC}, we recall relevant notions, results and the
aforementioned conjecture from~\cite{wollic12-paper}.  Section
\ref{section:pres-under-k-ary-covered-ext} introduces a generalization
of the classical notion of ``preservation under extensions'' and
formulates a dual version of the conjecture in terms of this notion.
In Section \ref{section:proof-of-conjecture}, we formally prove the
dual formulation of the conjecture, thereby proving the original
conjecture as well.  In Sections
\ref{section:finite-cores-and-finitary-covered-ext} and
\ref{section:generalizations}, we consider natural generalizations of
our notions and results, that yield a more general set of preservation
theorems.  In particular, we prove a generalization of the extensional
version of the {\lt} theorem. An interpolant-based approach to proving
the results proved till Section \ref{section:generalizations} is
presented in Section \ref{section:interpolants}.  In Sections
\ref{section:characterizing-Sigma_n-theories} and
\ref{section:theories-in-PSCf-and-PSC(k)-are-equivalent-to-Sigma_2-theories},
we present partial results towards generalizing the substructural
version of the {\lt} theorem -- in particular, we show that theories
that are preserved under substructures modulo $k$-sized cores are
equivalent to $\Sigma^0_2$ theories. We also show that the latter kind
of theories are more general than the former kind. In this process, we
prove a preservation theorem that provides a semantic characterization
of $\Sigma^0_n$ theories for each natural number $n$.  Finally, we
conclude in Section~\ref{section:conclusion} with some discussion and
directions for future work.

\section{Background}\label{section:PSC}

We assume that the reader is familiar with standard notation and
terminology used in the syntax and semantics of FO
(see~\cite{chang-keisler}). A \ti{vocabulary} $\tau$ is a set of
predicate, function and constant symbols.  In this paper, we restrict
ourselves to \emph{finite} vocabularies.  We denote by $FO(\tau)$ the
set of all FO formulae over vocabulary $\tau$.  A sequence $(x_1,
\ldots, x_k)$ of variables is denoted by $\bar{x}$. A $FO(\tau)$
formula $\psi$ having free variables $\bar{x}$ is denoted by
$\psi(\bar{x})$.  A formula with no free variables is called a
\emph{sentence}. A \emph{theory}, resp. \emph{theory over $\tau$}, is
a set of sentences, resp. $FO(\tau)$ sentences.  We abbreviate a block
of quantifiers of the form $Q x_1 \ldots Q x_k$ by $Q \bar{x}$, where
$Q \in \{\forall, \exists\}$.  We denote the natural numbers
\emph{including zero} by $\mathbb{N}$.  For every non-zero $k \in
\mathbb{N}$, we denote by $\Sigma^0_k$ (resp. $\Pi^0_k$), all FO
sentences in prenex normal form, whose quantifier prefix begins with a
$\exists$ (resp. $\forall$) and consists of $k-1$ alternations of
quantifiers.  We use the standard notions of $\tau$-structures,
substructures (denoted as $M \subseteq N$) and extensions, as defined
in ~\cite{chang-keisler}, and study preservation theorems over
\emph{arbitrary structures}.  By the \emph{size} (or \ti{power}) of a
structure $M$, we mean the cardinality of its universe, and denote it
by $|M|$.  A class of structures is called \emph{elementary}
(resp. \emph{basic elementary}) if it is definable by a FO theory
(resp. an FO sentence).  In this paper, we restrict ourselves to
definability by FO sentences until Section
\ref{section:generalizations}. Subsequently, we relax this restriction
to also include definability by formulae and theories.

We begin by recalling a generalization of the notion of ``preservation
under substructures'', introduced in~\cite{wollic12-paper}.
\begin{defn}[ref.~\cite{wollic12-paper}]\label{defn:PSC(k)}
For $k \in \mathbb{N}$, a class $S$ of structures is said to be
\emph{preserved under substructures modulo $k$-sized cores}, denoted
by $S \in {\PSC}(k)$, if for every structure $M \in S$, there exists
an at most $k$-sized subset $C$ of the universe of $M$ such that if $N
\subseteq M$ and $N$ contains $C$, then $N \in S$.  The set $C$ is
called a \emph{core of $M$ w.r.t. $S$}. If $S$ is clear from context,
we simply call $C$ a \emph{core of $M$}.
\end{defn}

An example of a class in ${\PSC}(0)$ (and hence in ${\PSC}(k)$ for
every $k \in \mathbb{N}$) is the class of all acyclic directed
graphs\footnote{A directed graph can be viewed as a $\tau$-structure
  where $\tau =\{E\}$ and $E$ is a binary predicate.}.  It is well
known that this class is not FO-definable.  Hence, ${\PSC}(k)$
contains classes not definable in FO for every $k \in \mathbb{N}$.
Let $PSC(k)$ denote the collection of FO-definable classes in
${\PSC}(k)$.  We identify classes in $PSC(k)$ with their defining FO
sentences.  As an example, for $k \in \mathbb{N}$, consider the class
$S_k$ of all graphs containing a $k$ length cycle as a subgraph. It is
clear that for any graph $G$ in $S_k$, the vertices of any cycle of
length $k$ in $G$ form a core of $G$.  Hence $S_k \in {\PSC}(k)$.
Since $S_k$ is definable using a $\Sigma^0_1$ sentence $\phi$ having
$k$ existential quantifiers, we say that the defining sentence $\phi$
is in $PSC(k)$.

Since ${\PSC}(0)$ coincides with the property of preservation under
substructures, we abbreviate ${\PSC}(0)$ as ${\PS}$ and $PSC(0)$ as
$PS$ in the following discussion.  The substructural version of the
{\lt} theorem for sentences can now be stated as follows.
\begin{theorem}[{\L}o{\'s}-Tarski]\label{theorem:los-tarski-primal}
A sentence is in $PS$ iff it is equivalent to a $\Pi^0_1$ sentence.
\end{theorem}
In attempting a syntactic characterization of $PSC(k)$ that
generalizes Theorem~\ref{theorem:los-tarski-primal}, the statement of
the following theorem was put forth as a conjecture in
~\cite{wollic12-paper}.  While it was shown to hold in several special
cases, it was not resolved in its entirety.  In this paper, we
formally prove the conjecture in its generality.
\begin{theorem}\label{theorem:PSC(k)=E^kA*}
A sentence is in $PSC(k)$ iff it is equivalent to a $\Sigma^0_2$
sentence with $k$ existential quantifiers.
\end{theorem}
It is easy to see that given a $\Sigma^0_2$ sentence
$\phi\equiv\exists x_1 \ldots \exists x_k \forall \bar{y}~\varphi(x_1,
\ldots, x_k, \bar{y})$ and a structure $M$ such that $M \models \phi$,
the {\em witnesses} $a_1, \ldots, a_k$ of $x_1, \ldots, x_k$ for which
$M \models \forall \bar{y}~\varphi(a_1,a_2,\ldots, a_k,\bar{y})$ form
a core of $M$. Therefore, $\phi \in PSC(k)$.  However, contrary to
intuition, witnesses and cores cannot always be equated!  For example,
consider the sentence $\phi\equiv \exists x \forall y E(x,y) \in
PSC(1)$ and the structure $M = (\mathbb{N}, \leq)$ (i.e. the natural
numbers with the usual ordering).  Clearly, $M \models \phi$ and the
only witness for $x$ is the minimum element $0 \in \mathbb{N}$.  In
contrast, every singleton subset of $\mathbb{N}$ forms a core of $M$!
This is because, $\mathbb{N}$ being well-ordered by $\leq$, for every
$x \in \mathbb{N}$, every substructure of $M$ containing $x$ has a
minimum element.  Therefore, there are many more cores than witnesses
in this example.

In light of Theorem~\ref{theorem:los-tarski-primal}, it is easy to see
that for relational vocabularies, if $\phi \in PS$, then $\phi$ must
be equivalent to a sentence of the form $\psi\equiv \forall y_1 \ldots
\forall y_l~ \phi|_{\{y_1, \ldots, y_l\}}$ for some $l \in
\mathbb{N}$, where $\phi|_{\{y_1, \ldots, y_l\}}$ is a quantifier-free
       {\em relativized} formula asserting that $\phi$ is true in the
       substructure induced by $y_1, \ldots, y_l$
       (see~\cite{wollic12-paper} for details).  In other words,
       checking the truth of $\phi$ (known to be in $PS$) in a
       structure $M$ amounts to checking its truth for all finite
       substructures of $M$ upto a suitably large size. In view of
       this, it is tempting to claim that if $\phi \in PSC(k)$, then
       checking the truth of $\phi$ in a structure $M$ amounts to
       finding a subset $C$ of $M$ of size at most $k$ and checking
       the truth of $\phi$ in all suitably large but finite
       substructures of $M$ that contain $C$.  However, this claim is
       incorrect.  To see why this is so, consider $\phi\equiv \exists
       x \forall y E(x,y)$ and the structure $M = (\mathbb{Z}, \leq)$
       (i.e., integers with the usual ordering).  Clearly $\phi \in
       PSC(1)$ and $M \not\models \phi$.  However, every {\em finite}
       substructure of $M$ has a minimum element, and hence models
       $\phi$!  This example illustrates a key difference between
       Theorem~\ref{theorem:los-tarski-primal} and
       Theorem~\ref{theorem:PSC(k)=E^kA*}.  Specifically, although
       Theorem~\ref{theorem:PSC(k)=E^kA*} asserts that every $\phi \in
       PSC(k)$ is equivalent to a sentence of the form
       $\exists^k\forall^* \varphi$ where $\varphi$ is
       quantifier-free, it reveals no information about the form of
       $\varphi$.

As defined in~\cite{wollic12-paper}, let ${\PSC} = \bigcup_{k \ge 0}
{\PSC}(k)$ and $PSC = \bigcup_{k \ge 0} PSC(k)$. Then Theorem
\ref{theorem:PSC(k)=E^kA*} yields the following corollary, which was
proven using other techniques in~\cite{wollic12-paper}.
\begin{corollary}[ref.~\cite{wollic12-paper}]\label{corollary:PSC=Sigma^0_2}
A sentence is in $PSC$ iff it is equivalent to a $\Sigma^0_2$
sentence.
\end{corollary}
The next two sections introduce a notion dual to that of preservation
under substructures modulo $k$-sized cores, formulate a dual version
of Theorem~\ref{theorem:PSC(k)=E^kA*} using this notion, and provide a
proof of the dual formulation.  A proof of
Theorem~\ref{theorem:PSC(k)=E^kA*} follows immediately from the dual
result.

\section{Preservation under $k$-ary Covered Extensions}\label{section:pres-under-k-ary-covered-ext}

The classical notion of ``extension of a structure'' can be naturally
generalized to \emph{extension of a collection of structures} as
follows.  A structure $M$ is said to be an extension of a collection
$R$ of structures if for each $N \in R$, we have $N \subseteq M$.
We now define a special kind of extensions of a collection of
structures.
\begin{defn}
For $k \in \mathbb{N}$, a structure $M$ is said to be a \emph{$k$-ary
  covered extension} of a non-empty collection $R$ of structures if
(i) $M$ is an extension of $R$, and (ii) for every subset $S$, of size
at most $k$, of the universe of $M$, there is a structure in $R$ that
contains $S$.  We call $R$ a \emph{$k$-ary cover} of $M$.
\end{defn}
As an example, let $M$ be a graph on $n$ vertices and let $R$ be the
collection of all $r$ sized induced subgraphs of $M$, where $1 \leq r
< n$. Then $M$ is a $k$-ary covered extension of $R$ for every $k$ in
$\{0, \ldots, r\}$.

Note that a $0$-ary covered extension of $R$ is simply an extension of
$R$.  For $k > 0$, the universe of a $k$-ary covered extension of $R$
is necessarily the union of the universes of the structures in $R$.
However, different $k$-ary extensions of $R$ can differ in the
interpretation of predicates (if any) of arity greater than $k$.  Note
also that a $k$-ary covered extension of $R$ is an $l$-ary covered
extension of $R$ for every $l \in \{0, \ldots, k\}$.

\begin{defn}
Given $k \in \mathbb{N}$, a class $S$ of structures is said to be
\emph{preserved under $k$-ary covered extensions}, denoted $S \in
{\PCE}(k)$, if for every collection $R$ of structures of $S$, if $M$ is
a $k$-ary covered extension of $R$, then $M \in S$.
\end{defn}
An example of a class in ${\PCE}(k)$ is the class $S_k$ of all graphs
not containing a cycle of length $k$.  Let $M$ be a $k$-ary covered
extension of $R$, where $R$ is a collection of structures of $S_k$.
It is easy to see that $M$ is also in $S_k$.  For if not, $M$ must
contain a cycle of length $k$.  As $R$ is a $k$-ary cover of $M$, this
cycle must be contained in some $N \in R$.  This contradicts the fact
that $N \in S_k$.

Let the collection of FO definable classes in ${\PCE}(k)$ be denoted
by $PCE(k)$.  As before, we identify classes in $PCE(k)$ with their
defining FO sentences.  It is easy to see that if $l, k \in
\mathbb{N}$ and if $l \leq k$, then ${\PCE}(l) \subseteq {\PCE}(k)$
and $PCE(l) \subseteq PCE(k)$.  Furthermore, the heirarchies within
${\PCE}$ and $PCE$ are strict.  Consider $\phi \in PCE(k)$ over the
empty vocabulary given by $\phi = \forall x_1 \ldots \forall x_k $
$\bigvee_{1 \leq i < j \leq k} (x_i = x_j)$. The sentence $\phi$
asserts that there are strictly fewer than $k$ elements in any
model. It is easy to see that $\phi \in PCE(k)$. To see that $\phi
\notin PCE(l)$ for $l < k$, consider $M$ containing exactly $k$
elements in its universe. Clearly $M \not\models \phi$. Consider the
collection $R$ of all $l$-sized substructures of $M$. It is easy to
check that $R$ is a $l$-ary cover for $M$. However each structure in
$R$ models $\phi$. Then $\phi \notin PCE(l)$ for $l < k$. This shows
the strict heirarchy within $PCE$.  The above argument also shows that
${\PCE}(l)$ is strictly contained in ${\PCE}(k)$ -- $\phi$ witnesses
this strict inclusion.

The classical notion of preservation under extensions is easily seen
to coincide with ${\PCE}(0)$.  This is because a class of structures
$S$ is preserved under extensions iff it is preserved under extensions
of any collection $R$ of structures in $S$.  This motivates
abbreviating ${\PCE}(0)$ as ${\PE}$ and $PCE(0)$ as $PE$ in the
subsequent discussion.  Analogous to the definitions of ${\PSC}$ and
$PSC$, we define ${\PCE} = \bigcup_{k \ge 0} {\PCE}(k)$ and $PCE =
\bigcup_{k \ge 0} PCE(k)$.  The strictness of the ${\PCE}(k)$ and
$PCE(k)$ hierarchies imply that ${\PCE}$ and $PCE$ strictly generalize
${\PE}$ and $PE$ respectively.

With the above notation, the extensional version of the {\lt} theorem
for sentences can be stated as follows.
\begin{theorem}[\lt]\label{theorem:los-tarski-dual}
A sentence is in $PE$ iff it is equivalent to a $\Sigma^0_1$ sentence.
\end{theorem}

The duality between ${\PSC}(k)$ and ${\PCE}(k)$ is formalized by
the following lemma.
\begin{lemma}[${\PSC}(k)$-${\PCE}(k)$ duality]\label{lemma:PSC(k)-PCE(k)-duality-for-any-class-of-structures}
A class $S$ of structures is in ${\PSC}(k)$ iff its complement
$\overline{S}$ is in ${\PCE}(k)$.
\end{lemma}
\ti{Proof}: \underline{If:}~ Suppose $\overline{S} \in {\PCE}(k)$ but
$S \notin {\PSC}(k)$. Then there exists $M \in S$ s.t. for every set
$A$ of at most $k$ elements from $M$, there is a substructure $N_A$ of
$M$ that contains $A$ but is not in $S$.  In other words, $N_A \in
\overline{S}$. Then $R = \{N_A \mid A ~\text{is a subset, of size }$
$\text{at most }k, \text{of } M \}$ is a $k$-ary cover of $M$. Since
$N_A \in \overline{S}$ for all $N_A \in R$ and since $\overline{S} \in
         {\PCE}(k)$, it follows that $M \in \overline{S}$ -- a
         contradiction.

\underline{Only If}:~ Suppose $S \in {\PSC}(k)$ but $\overline{S}
\notin {\PCE}(k)$. Then there exists $M \in S$ and a $k$-ary cover $R$
of $M$ such that every structure $N \in R$ belongs to
$\overline{S}$. Since $M \in S$, there exists a core $C$ of $M$
w.r.t. $S$ of size at most $k$. Consider the structure $N_C \in R$
that contains $C$ -- this exists since $R$ is a $k$-ary cover of
$M$. Then $N_C \in S$ since $C$ is a core of $M$ -- a
contradiction.\epf
\begin{corollary}\label{corollary:PSC(k)-PCE(k)-duality-for-sentences}
Let $\phi$ be a FO sentence. Then $\phi \in PSC(k)$ iff $\neg \phi
\in PCE(k)$.
\end{corollary}
As seen earlier, all $\Sigma^0_2$ sentences with $k$ existential
quantifiers are in $PSC(k)$.  By Corollary
\ref{corollary:PSC(k)-PCE(k)-duality-for-sentences}, all $\Pi^0_2$
sentences with $k$ universal quantifiers are in $PCE(k)$.  In the next
section, we show that the converse is also true, yielding the
following theorem.
\begin{theorem}\label{theorem:PCE(k)=A^kE*}
A sentence is in $PCE(k)$ iff it is equivalent to a $\Pi^0_2$ sentence
with $k$ universal quantifiers.
\end{theorem}
Theorem \ref{theorem:PCE(k)=A^kE*} and Corollary
\ref{corollary:PSC(k)-PCE(k)-duality-for-sentences} together prove
Theorem \ref{theorem:PSC(k)=E^kA*}.
Theorem~\ref{theorem:PCE(k)=A^kE*} also yields a new characterization
of the $\Pi^0_2$ fragment of FO, as given by the following corollary.
\begin{corollary}\label{corollary:PCE=A*E*}
A sentence is in $PCE$ iff it is equivalent to a $\Pi^0_2$ sentence.
\end{corollary}

\section{Proof of Theorem \ref{theorem:PCE(k)=A^kE*}}\label{section:proof-of-conjecture}

We begin by recalling some notions from classical model
theory~\cite{chang-keisler}.  Given a vocabulary $\tau$ and a cardinal
$\alpha$, let $\tau_\alpha$ be the vocabulary obtained by expanding
$\tau$ with fresh constants $c_1, \ldots, c_\alpha$.  Given a
$\tau$-structure $M$, the theory of $M$, denoted $Th(M)$, is the set
of all $FO(\tau)$ sentences true in $M$.  Given $\tau$-structures $M$
and $N$, we call $M$ an \emph{elementary extension} of $N$ if for all
$k \in \mathbb{N}$, for all $FO(\tau)$ formulae $\varphi(x_1, \ldots,
x_k)$ and for all $k$-tuples $\bar{a}$ from $N$, $M \models
\varphi(\bar{a}) \leftrightarrow N \models \varphi(\bar{a})$.  A
$\tau$-type $\Sigma(v)$ is a maximally consistent set of
$\tau$-formulae having a single free variable $v$.  In other words,
for every $\tau$-formula $\psi(v)$, exactly one of $\psi$ or $\neg
\psi$ belongs to $\Sigma(v)$. A structure $M$ is said to
\emph{realize} the type $\Sigma(v)$ if there is an element $a$ of $M$
such that $M \models \Sigma(a)$. Finally, we recall the notion of
\emph{saturation}, which is crucially used in our proof.

\begin{defn}[Saturation, ref.~\cite{chang-keisler}]
Given a cardinal $\lambda$, a $\tau$-structure $M$ is said to be
$\lambda$-saturated if for every subset $X = \{b_1, \ldots,
b_{\alpha}\}$ of the universe of $M$ such that $\alpha < \lambda$, the
$\tau_\alpha$ expansion $(M, b_1, \ldots, b_\alpha)$ realizes every
$\tau_\alpha$-type $\Sigma(v)$ that is consistent with $Th(M, b_1,
\ldots, b_\alpha)$.
\end{defn}

Our proof makes use of the following results from Chapter 5 of
~\cite{chang-keisler}.

\begin{proposition}[ref.~\cite{chang-keisler}]\label{proposition:extending-saturations-to-expansions}
Given an infinite cardinal $\lambda$ and a $\lambda$-saturated
$\tau$-structure $M$, for every $k$-tuple $(a_1, \ldots, a_k)$ of
elements from $M$ where $k \in \mathbb{N}$, the $\tau_k$ structure
$(M, a_1, \ldots, a_k)$ is also $\lambda$-saturated.
\end{proposition}

\begin{proposition}[ref.~\cite{chang-keisler}]\label{proposotion:finite-structures-are-saturated}
A $\tau$-structure is finite iff it is $\lambda$-saturated for
all cardinals $\lambda$.
\end{proposition}

\begin{theorem}[ref.~\cite{chang-keisler}]\label{theorem:existence-of-saturated-elementary-extensions}
Let $\tau$ be a finite vocabulary, $\lambda$ be an infinite cardinal
and $M$ be a $\tau$-structure such that $\omega \leq |M| \leq
2^{\lambda}$ ($|M|$ denotes the power of $M$). Then there is a
$\lambda$-saturated elementary extension of $M$ of power $2^\lambda$.
\end{theorem}

\begin{theorem}[ref.~\cite{chang-keisler}]\label{theorem:embedding-conditions}
Given $\tau$-structures $M$ and $N$ and a cardinal $\lambda$, suppose
that (i) $M$ is $\lambda$-saturated, (ii) $\lambda \ge |N|$, and (iii)
every existential sentence true in $N$ is also true in $M$. Then $N$
is isomorphically embeddable in $M$.
\end{theorem}

Putting Theorem
\ref{theorem:existence-of-saturated-elementary-extensions} and
Proposition \ref{proposotion:finite-structures-are-saturated} together,
we get the following.
\begin{corollary}\label{corollary:always-existence-of-saturated-elementary-extensions}
Let $\tau$ be a finite vocabulary. For every $\tau$-structure $M$,
there exists a $\beta$-saturated elementary extension of $M$ for some
cardinal $\beta \ge \omega$.
\end{corollary}

Towards our syntactic characterization of $PCE(k)$, we first prove the
following.

\begin{lemma}\label{lemma:k-ary-covers-for-saturated-models}
Let $\tau$ be a finite vocabulary and let $T$ be a consistent theory
over $\tau$.  Let $\Gamma$ be the set of all $\forall^k \exists^*$
consequences of $T$.  Then for all infinite cardinals $\lambda$, for
every $\lambda$-saturated structure $M$, if $M \models \Gamma$, then
there exists a $k$-ary cover $R$ of $M$ such that $N \models T$ for
every $N \in R$.
\end{lemma}
\ti{Proof}: We show that for every subset $S$, of size at most $k$, of
the universe of $M$, there is a substructure $M_s$ of $M$ containing
$S$ such that $M_s \models T$. Then the set $K = \{M_s \mid S
~\text{is a subset, of size at most}~ k,\text{ of the universe of~}
M\}$ forms the desired $k$-ary cover of $M$.  Let $S = \{a_1, \ldots,
a_r\}$ be a subset of the universe of $M$, where $r \le k$. To show
the existence of $M_s$, it suffices to show that there exists a
$\tau_r$-structure $N$ such that (i) $|N| \le \lambda$, (ii) every
$\exists^*$ sentence true in $N$ is also true in $(M, a_1, \ldots,
a_r)$, and (iii) $N \models T$.  Since $M$ is $\lambda$-saturated, by
Proposition \ref{proposition:extending-saturations-to-expansions},
$(M, a_1, \ldots, a_r)$ is also $\lambda$-saturated. Then, from
Theorem \ref{theorem:embedding-conditions}, $N$ is isomorphically
embeddable into $(M, a_1, \ldots, a_r)$. Then the $\tau$-reduct of the
copy of $N$ in $(M, a_1, \ldots, a_r)$ can serve as $M_s$.  The proof
is therefore completed by showing the existence of $N$ with the above
properties.

Let $P$ be the set of all $\forall^*$ sentences of $FO(\tau_r)$ that
are true in $(M, a_1, \ldots, a_r)$.  Suppose $Z = T \cup P$ is
inconsistent. By the compactness theorem, there is a finite subset of
$Z$ that is inconsistent. Since $P$ is closed under taking finite
conjunctions and since each of $P$ and $T$ is consistent, there is a
sentence $\psi$ in $P$ such that $T \cup \{\psi\}$ is inconsistent. In
other words, $T \rightarrow \neg \psi$.  Since $T$ is a theory over
$\tau$ and $\psi$ is a sentence over $\tau_r$, by
$\forall$-introduction, we have $T \rightarrow \varphi$, where
$\varphi \equiv \forall x_1 \ldots \forall x_r \neg \psi[ c_1 \mapsto
  x_1; \ldots; c_r \mapsto x_r]$, the variables $x_1, \ldots, x_r$ are
fresh, and $c_i \mapsto x_i$ denotes substitution of $c_i$ by $x_i$.
Since $\neg \psi$ is a $\exists^*$ sentence over $\tau_r$, $\varphi$
is a $\forall^r \exists^*$ sentence over $\tau$.  Since $r \le k$,
$\varphi$ can be seen as a $\forall^k \exists^*$ sentence (by
introducing redundant $\forall$s if $r < k$). By the definition of
$\Gamma$, we must have $\varphi \in \Gamma$, and hence $M \models
\varphi$. In other words, $(M, a_1, \ldots, a_r) \models \neg
\psi$. This contradicts the fact that $\psi \in P$.  Therefore, $Z$
must be consistent.  By L\"owenheim-Skolem theorem, there is a model
$N$ of $Z$ of power at most $\lambda$.  Since $N$ models every
$\forall^*$ sentence true in $(M, a_1, \ldots, a_r)$, every
$\exists^*$ sentence true in $N$ must be true in $(M, a_1, \ldots,
a_r)$. Finally, since $N \models T$, $N$ is indeed as desired.  \epf\\

We complete the proof of Theorem \ref{theorem:PCE(k)=A^kE*} now. If
$\phi$ is unsatisfiable, we are done. Otherwise, let $\Gamma$ be the
set of all $\forall^k \exists^*$ consequences of $\phi$.  Clearly,
$\phi \rightarrow \Gamma$.  For the converse, suppose $M \models
\Gamma$. By Corollary
\ref{corollary:always-existence-of-saturated-elementary-extensions},
there is a $\beta-$saturated elementary extension $M^+$ of $M$ for
some $\beta \ge \omega$.  Then $M^+ \models \Gamma$. Taking $T =
\{\phi\}$, by Lemma \ref{lemma:k-ary-covers-for-saturated-models},
there exists a $k$-ary cover $R$ of $M^+$ such that for every $N \in
R$, $N \models T$ i.e.  $N \models \phi$.  Since $\phi \in PCE(k)$, it
follows that $M^+ \models \phi$.  As $M^+$ and $M$ are elementarily
equivalent, $M \models \phi$ and hence $\Gamma \rightarrow \phi$.
This shows that $\phi \leftrightarrow \Gamma$. By the compactness
theorem, $\phi$ is equivalent to a finite conjunction of sentences of
$\Gamma$. Since $\Gamma$ is closed under finite conjunctions, $\phi$
is equivalent to a $\forall^k \exists^*$ sentence. \epf\\

We remark that the above proof goes through over any class of
structures satisfying the compactness theorem. As a special case then,
Theorem \ref{theorem:PCE(k)=A^kE*} and (hence) Theorem
\ref{theorem:PSC(k)=E^kA*} are true \emph{modulo} theories. Thus, we
get a complete subsumption of the {\lt} theorem for sentences in both
its senses, primal and dual.

\section{Finite Cores and Finitary Covers}\label{section:finite-cores-and-finitary-covered-ext}

We recall from ~\cite{wollic12-paper} the following notion which
generalizes ${\PSC}(k)$.

\begin{defn}[ref. ~\cite{wollic12-paper}]

A class $S$ of structures is said to be \emph{preserved under
  substructures modulo a finite core}, denoted $S \in \PSCf$, if for
every structure $M \in S$, there exists a finite subset $C$ of the
universe of $M$ such that if $N \subseteq M$ and $N$ contains $C$,
then $N \in S$. The set $C$ is called a \emph{core of $M$ w.r.t. $S$}.
\end{defn}

As an example, the class $S$ of graphs containing cycles is in
{\PSCf}. Since {\PSCf} contains classes like $S$ that are not FO
definable, we let $PSC_f$ denote the collection of FO definable
classes in {\PSCf}. We identify classes in $PSC_f$ with their defining
FO sentences. The following results were then proved.

\begin{theorem}[ref.~\cite{wollic12-paper}]\label{theorem:PSC_f=Sigma^0_2}
A sentence is in $PSC_f$ iff it is equivalent to a $\Sigma^0_2$
sentence.
\end{theorem}

\begin{corollary}[ref.~\cite{wollic12-paper}]\label{corollary:PSCf=PSC}
$PSC_f = PSC$.
\end{corollary}

\begin{lemma}[ref.~\cite{wollic12-paper}]\label{lemma:non-recursiveness-for-finite-core=bounded-core}
For every recursive function $f: \mathbb{N} \rightarrow \mathbb{N}$,
there is a sentence $\phi \in PSC_f$ which is not in $PSC(k)$ for any
$k \leq f(|\phi|)$.
\end{lemma}

We now give analogous notions and results in the dual setting.  We
first define the notion of a \emph{finitary covered extension} which
is a natural generalization of the notion of a $k$-ary covered
extension introduced in Section
\ref{section:pres-under-k-ary-covered-ext}.

\begin{defn}
A structure $M$ is said to be a \emph{finitary covered extension} of a
collection $R$ of structures if (i) $M$ is an extension of $R$ (ii)
for every \emph{finite} subset $S$ of the universe of $M$, there is a
structure in $R$ containing $S$. We then call $R$ as a \emph{finitary
  cover} of $M$.
\end{defn}

If $M$ is a finitary covered extension of $R$, then $R$ is necessarily
non-empty. Further, $M$ is the unique finitary covered extension of
$R$ since all predicates and function symbols have finite
arity. Finally, $M$ is also a $k$-ary covered extension of $R$ for all
$k \in \mathbb{N}$.

\begin{defn}
A class $S$ of structures is said to be \emph{preserved under finitary
  covered extensions}, denoted $S \in \PCEf$, if for every collection
$R$ of structures of $S$, if $M$ is a finitary covered extension of
$R$, then $M \in S$.
\end{defn}

It is easy to see that ${\PCE} \subseteq {\PCEf}$. Since ${\PCE}$
contains non-FO definable classes, these are also in {\PCEf} and hence
let $PCE_f$ denote the collection of FO definable classes in
${\PCEf}$. As usual, we identify classes in $PCE_f$ with their
defining FO sentences. We now have the following duality result
similar to Lemma
\ref{lemma:PSC(k)-PCE(k)-duality-for-any-class-of-structures}. The
proof is exactly like the proof of the latter -- just replace `$k$' in
the latter proof by `finite'.

\begin{lemma}[${\PSCf}$-${\PCEf}$ duality]\label{lemma:PSCf-PCEf-duality-for-any-class-of-structures}
A class $S$ of structures is in ${\PSCf}$ iff its complement
$\overline{S}$ is in ${\PCEf}$. In particular, if $\phi$ is a FO
sentence, then $\phi \in PSC_f$ iff $\neg \phi \in PCE_f$.
\end{lemma}

Lemma \ref{lemma:PSCf-PCEf-duality-for-any-class-of-structures} and
Theorem \ref{theorem:PSC_f=Sigma^0_2} give us the following
characterization of $PCE_f$.

\begin{theorem}\label{theorem:PCE_f=Pi^0_2}
A sentence is in $PCE_f$ iff it is equivalent to a $\Pi^0_2$ sentence.
\end{theorem}

We remark that this result has an alternate proof very similar to that
of Theorem \ref{theorem:PCE(k)=A^kE*}. We now get a result analogous
to Corollary \ref{corollary:PSCf=PSC}.

\begin{corollary}\label{corollary:PCEf=PCE}
$PCE_f = PCE$.
\end{corollary}

Finally, we have Lemma
\ref{lemma:non-recursiveness-for-finite-cover=bounded-cover} below
analogous to Lemma
\ref{lemma:non-recursiveness-for-finite-core=bounded-core}. 

\begin{lemma}\label{lemma:non-recursiveness-for-finite-cover=bounded-cover}
For every recursive function $f: \mathbb{N} \rightarrow \mathbb{N}$,
there is a sentence $\phi \in PCE_f$ but which is not in $PCE(k)$ for
any $k \leq f(|\phi|)$.
\end{lemma}

\ti{Proof}: Suppose there is a recursive function $f: \mathbb{N}
\rightarrow \mathbb{N}$ such that if $\phi \in PCE_f$, then $\phi \in
PCE(k)$ for some $k \leq f(|\phi|)$. Then consider the function $g:
\mathbb{N} \rightarrow \mathbb{N}$ given by $g(n) = f(n + 1)$. Clearly
$g$ is also a recursive function. Now let $\phi$ be a sentence in $
PSC_f$. We will show that $\phi \in PSC(k)$ for some $k \leq
g(|\phi|)$. This would contradict Lemma
\ref{lemma:non-recursiveness-for-finite-core=bounded-core} and
complete our proof.

Since $\phi \in PSC_f$, by Lemma
\ref{lemma:PSCf-PCEf-duality-for-any-class-of-structures}, it follows
that $\neg \phi \in PCE_f$. Then by our assumption above, $\neg \phi
\in PCE(k)$ for some $k \leq f(|\neg \phi|) = f(1 + |\phi|) =
g(|\phi|)$. Then by Corollary
\ref{corollary:PSC(k)-PCE(k)-duality-for-sentences}, we have
$\phi \in PSC(k)$ for some $k \leq g(|\phi|)$. \epf.

\section{Generalizations to formulae and theories}\label{section:generalizations}
In this section, we generalize the semantic classes $PSC$, $PCE$ and
various subclasses of these seen earlier.  This is done by relaxing
the meaning of FO definability to include definability by theories (as
opposed to definability by FO sentences used so far).  Specifically,
the classes $PSC_f, PSC, PSC(k), PCE_f, PCE$ and $PCE(k)$ are now
(re-)defined to be subclasses of ${\PSCf}, {\PSC}, {\PSC}(k), {\PCEf},
{\PCE}$ and ${\PCE}(k)$, respectively, that are definable by FO
theories.  While a theory is conventionally a set of sentences, we
define a \emph{theory with free variables} $\bar{x}$ to be a set of FO
formulae, each of which has free variables $\bar{x}$.  Let
$T(\bar{x})$ be a $\tau$-theory with free variables $\bar{x}$, and let
$T'$ be the $\tau_n$-theory obtained by replacing each $x_i$ with
$c_i$ for $i \in \{1, \ldots, n\}$, where $n = |\bar{x}|$ and the
$c_i$'s are fresh constant symbols.  For each class $\mathcal{C}$ in
$\{PSC_f, PSC, PSC(k), PCE_f,$ $ PCE, PCE(k)\}$, we say that $T \in
\mathcal{C}$ iff $T' \in \mathcal{C}$.  The above generalizations of
semantic classes lead to natural generalizations of the preservation
theorems seen earlier.

\begin{theorem}\label{theorem:generalizations}
Let $T(\bar{x})$ be a theory with free variables $\bar{x}$.
\begin{enumerate}
\item \emph{(Gen. of Theorem \ref{theorem:PCE_f=Pi^0_2})}~~
  $T(\bar{x}) \in PCE_f$ iff $T(\bar{x})$ is equivalent to a theory of
  $\Pi^0_2$ formulae with free variables
  $\bar{x}$. \label{theorem:PCE_f=Pi^0_2-for-theories-with-free-vars}
\item \emph{(Gen. of Theorem \ref{theorem:PCE(k)=A^kE*})}~~
  $T(\bar{x}) \in PCE(k)$ iff $T(\bar{x})$ is equivalent to a theory
  of $\Pi^0_2$ formulae, each having free variables $\bar{x}$ and $k$
  universal quantifiers. \label{theorem:PCE(k)=A^kE*-for-theories-with-free-vars}
\end{enumerate}
\end{theorem}
The proof of each part of Theorem~\ref{theorem:generalizations} is
obtained by a straightforward adaptation of the proof of the
corresponding theorem it generalizes.  Putting $k = 0$ in Theorem
\ref{theorem:generalizations}(\ref{theorem:PCE(k)=A^kE*-for-theories-with-free-vars}),
we obtain the extensional version of the {\lt} theorem for theories
with free variables.

The proof of equivalence of $PCE$ and $PCE_f$ for FO sentences (see
Corollary~\ref{corollary:PCEf=PCE}) can be easily adapted to work for
FO formulae as well.  However, as Lemma
\ref{lemma:PCE_f-neq-PCE-for-theories} shows, $PCE_f$ strictly
subsumes $PCE$ for FO theories.

\begin{lemma}\label{lemma:PCE_f-neq-PCE-for-theories}
$PCE \subsetneq PCE_f$ for FO theories.
\end{lemma}
\ti{Proof}: That $PCE \subseteq PCE_f$ for theories is obvious.  To
prove the lemma, we present a theory of $\Pi^0_1$ sentences that is
not equivalent to any theory of $\forall^k \exists^*$ sentences, for
any $k \in \mathbb{N}$.  Let $T$ be a $\Pi^0_1$ theory over graphs
that asserts that there is no cycle of length $k$ for any $k \in
\mathbb{N}$.  The theory $T$ defines the class $S$ of all acyclic
graphs. If $T$ is equivalent to a theory of $\forall^k \exists^*$
sentences for some $k \in \mathbb{N}$, then by
Theorem~\ref{theorem:generalizations}(\ref{theorem:PCE(k)=A^kE*-for-theories-with-free-vars}),
$S$ must be in $PCE(k)$.  Hence, by
Lemma~\ref{lemma:PSC(k)-PCE(k)-duality-for-any-class-of-structures},
$\overline{S}$ (the complement of $S$) must be in ${\PSC}(k)$.  Now
consider a cycle $G$ of length $k+1$.  Clearly, $G \in \overline{S}$,
although every proper substructure of $G$ is in $S$.  This contradicts
that $\overline{S} \in {\PSC}(k)$.\epf\\

In contrast to the syntactic characterizations for theories in $PCE$
and $PCE_f$, we do not yet have syntactic characterizations for
theories in $PSC(k)$ and $PSC_f$.  The proof of Theorem
\ref{theorem:PSC(k)=E^kA*} for sentences can be easily adapted to work
for individual formulae in $PSC(k)$.  Recall that this proof proceeds
by characterizing the negations of sentences in $PSC(k)$.
Unfortunately, the same approach cannot be taken for characterizing
theories in $PSC(k)$ because negations of theories need not be
theories, as implied by the following result.

\begin{proposition}[ref.~\cite{chang-keisler}]
Suppose $S$ and $\overline{S}$ are both elementary classes of
structures. Then $S$ and $\overline{S}$ are both basic elementary.
\end{proposition}
Thus, if a theory $T$ in $PSC(k)$ defines a class $S$ of structures,
then while it is certain that $\overline{S} \in \PCE(k)$, it need not
be that $\overline{S} \in PCE(k)$. Hence $\overline{S}$ need not be
definable by a theory.  In which case, the characterization for
theories in $PCE(k)$ cannot be applied.

A natural proposal for characterizing theories $T$ in $PSC(k)$ is to
assert that $T$ is equivalent to a theory of $\exists^k \forall^*$
sentences. However, consider the theory $T = \{ \exists x E(x, x),$ $
\exists y \neg E(y, y)\}$ and a graph $G = (V, E) = (\{a, b\}, \{(a,
a)\})$. It is easy to check that $G$ cannot have a core of size
1. Thus, $T \notin PSC(1)$, and the proposal is falsified.  A modified
proposal asserts that $T \in PSC(k)$ iff $T$ is equivalent to an
infinitary logic sentence of the form $\exists^k \bar{x} \bigwedge_{i
  \in I} \psi_i(\bar{x})$, where $I$ is a set of indices and $\psi_i$
is a $\Pi^0_1$ formula with free variables $\bar{x}$, for each $i \in
I$. It is easy to check that any sentence of this form is indeed in
$PSC(k)$.  However, we have not yet been able to prove the converse.
In the next section, we suggest a possible approach to the problem of
characterizing theories in $PSC(k)$ and $PSC_f$.

\section{Characterizations using Interpolants}\label{section:interpolants}

Given a theory $T \in PS$, an interesting question is: Does there
exist a finite subset of $T$ which also is in $PS$? The following is a
recent unpublished result by Parikh that he proved in our discussions
with him (see~\cite{parikh-PS} for a proof).

\begin{theorem}[Parikh]\label{theorem:parikh-PS}
There is a theory in $PS$ s.t. no finite subset of it is in $PS$.
\end{theorem}

Given a theory $T \in PS$, call a finite subset $S$ of $T$ as
\emph{well-behaved} if $S$ is in $PS$. Then Theorem
\ref{theorem:parikh-PS} shows that there may not exist any
well-behaved finite subset of $T$.  We will however show below that
for each finite subset $S$ of $T$, there always exists, an
\emph{underapproximation} of it implied by $T$ which is
well-behaved. In other words, there is a sentence $\xi_S$ such that
(i) $T \rightarrow \xi_S$ and $\xi_S \rightarrow S$ (ii) $\xi_S \in
PS$.  Towards this, we recall from the literature~\cite{hodges} that,
given theories $Z$ and $T$ and a sentence $\psi$, a sentence $\xi$ is
said to be an \emph{interpolant between $T$ and $\psi$ modulo $Z$} if
$Z \vdash (T \rightarrow \xi \wedge \xi \rightarrow \psi)$. If $Z$ is
empty, then $\xi$ is simply called an \emph{interpolant between $T$
  and $\psi$}. The following result is a simple generalization of a
result from ~\cite{hodges}. There, the result below appears with a
sentence instead of $T$.

\begin{proposition}\label{proposition:generalizing-hodges-interpolant-result}
Let $Z, T$ be theories and $\psi$ be a sentence. The following are
equivalent:
\begin{enumerate}
\item If $M$ and $N$ model $Z$, $N \subseteq M$ and $M \models T$,
  then $N \models \psi$.
\item There is a $\Pi^0_1$ interpolant $\xi$ between $T$ and $\psi$
  modulo $Z$.
\end{enumerate}
\end{proposition}

We adopt two conventions in what follows: (a) $S \subseteq_f T$
denotes that $S$ is a finite subset of $T$. (b) We identify a finite
theory with the (finite) conjunction of the sentences in it. Now,
taking $Z$ to be empty and $\psi$ to be $S$, one sees that condition
(1) of Proposition
\ref{proposition:generalizing-hodges-interpolant-result} is true,
yielding the following.

\begin{corollary}\label{corollary:PS-interpolant}
Given a theory $T \in PS$ and $S \subseteq_f T$, there exists a
$\Pi^0_1$ interpolant $\xi_S$ between $T$ and $S$.
\end{corollary}

Indeed $\xi_S$ is the well-behaved underapproximation of $S$ we had
referred to above. Interestingly then, for a theory in $PS$, while the
{\lt} theorem for theories states the existence of a universal theory
equivalent to $T$, we can actually \emph{construct} one such theory
using the interpolants guaranteed by Corollary
\ref{corollary:PS-interpolant}. 

\begin{theorem}\label{theorem:interpolant-perspective-of-los-tarski-theorem}
Let $T$ be theory in $PS$ and $\xi_S$ be a $\Pi^0_1$ interpolant
between $T$ and $S$ for $S \subseteq_f T$. Then $T$ is equivalent to
$V = \{\xi_S \mid S \subseteq_f T\}$.
\end{theorem}
\ti{Proof}: For each $S \subseteq_f T$, we have $T \rightarrow \xi_S$;
then $T \rightarrow V$. Further since $\xi_S \rightarrow S$, we have
$V \rightarrow W$ where $W = \bigwedge_{S \subseteq_f T}
\bigwedge_{\phi \in S} \phi$. But check that $W \leftrightarrow
T$.\epf\\

We now present a result similar to Proposition
\ref{proposition:generalizing-hodges-interpolant-result}, which can be
seen as a generalization of a dual version of the proposition. The
proof below uses ideas similar to those used in proving Theorem
\ref{theorem:PCE(k)=A^kE*}.

\begin{proposition}\label{proposition:hodges-type-interpolant-result-for-PCE(k)}
Let $Z, T$ be theories and $\psi$ be a sentence. Then given $k \in
\mathbb{N}$, the following are equivalent:
\begin{enumerate}
\item Let $R$ be a $k$-ary cover ~(resp. finitary cover)~ of
  $M$. Suppose $M \models Z$ and for each $N \in R$, $N \models Z \cup
  T$. Then $M \models \psi$.
\item There is a $\forall^k \exists^*$ ~(resp. $\Pi^0_2$)~ interpolant
  $\xi$ between $T$ and $\psi$ modulo $Z$.
\end{enumerate}
\end{proposition}
\ti{Proof}: We give the proof for $k$-ary covers only. The proof for
finitary covers is analogous. Also the proofs modulo theories $Z$ are
analogous to the proof below which considers $Z$ as empty.

\underline{$(2) \rightarrow (1)$}: Since $N \models T$ for each $N \in
R$ and since $\xi$ is an interpolant between $T$ and $\psi$, we have
that $N \models \xi$ for each $N \in R$. Since $\xi$ is a $\forall^k
\exists^*$ sentence, $\xi \in PCE(k)$. Then $M \models \xi$, as $R$ is
a $k$-ary cover of $M$. Finally since $\xi$ is an interpolant between
$T$ and $\psi$, $M \models \psi$.

\underline{$(1) \rightarrow (2)$}: Let $\Gamma$ be the set of all
$\forall^k \exists^*$ consequences of $T$. We will show that $\Gamma
\rightarrow \psi$. Then by the compactness theorem, a finite
conjunction of sentences in $\Gamma$ would imply $\psi$. Since
$\Gamma$ is closed under finite conjunctions, we would get a single
$\forall^k \exists^*$ sentence in $\Gamma$ which would imply
$\psi$. We can take this sentence to be the desired interpolant $\xi$.

Suppose $M \models \Gamma$. By Corollary
\ref{corollary:always-existence-of-saturated-elementary-extensions},
there is a $\beta-$saturated elementary extension $M^+$ of $M$ for
some $\beta \ge \omega$.  Then $M^+ \models \Gamma$. By Lemma
\ref{lemma:k-ary-covers-for-saturated-models}, there exists a $k$-ary
cover $R$ of $M^+$ such that $N \models T$ for every $N \in R$. Then
by the premises as stated in (1), $M^+ \models \psi$.  As $M^+$ and
$M$ are elementarily equivalent, $M \models \psi$. This shows that
$\Gamma \rightarrow \psi$. \epf

\begin{corollary}\label{corollary:PCE(k)-interpolant}
Given $k \in \mathbb{N}$, a theory $T \in PCE(k)$ ~(resp. $T \in
PCE_f$)~ and $S \subseteq_f T$, there exists a $\forall^k \exists^*$
(resp. $\Pi^0_2$) interpolant $\xi_S$ between $T$ and $S$.
\end{corollary}

This result gives the following alternate proof of Theorem
\ref{theorem:generalizations} for theories.

\begin{theorem}\label{theorem:proof-of-PCE(k)=A^kE*-via-interpolants}
Given a theory $T \in PCE(k)$ ~(resp. $T \in PCE_f$)~, let $\xi_S$ be
a $\forall^k \exists^*$ ~(resp. $\Pi^0_2$)~ interpolant between $T$
and $S$ for $S \subseteq_f T$. Then $T$ is equivalent to $\{\xi_S \mid
S \subseteq_f T\}$.
\end{theorem}

A natural question to ask in view of the above results is: Given a
theory $T$ in $PSC(k)$ does there exist a $\exists^k \forall^*$
interpolant between $T$ and $S$ for $S \subseteq_f T$? We have no
answer to this question yet. If this is true, then analogous to
Theorem \ref{theorem:proof-of-PCE(k)=A^kE*-via-interpolants}, we would
have that if $T \in PSC(k)$, then $T$ is equivalent to a theory of
$\exists^k \forall^*$ sentences.  For theories $T$ in $PSC(k)$, one
can even ask the following weaker question: Given $S \subseteq_f T$,
does there exist a $\Sigma^0_2$ interpolant $\xi_S$ between $T$ and
$S$?  If yes, then analogous to Theorem
\ref{theorem:proof-of-PCE(k)=A^kE*-via-interpolants}, $T$ would be
equivalent to a theory of $\Sigma^0_2$ sentences. Observe that if $T$
were equivalent to a theory of $\Sigma^0_2$ sentences, then for any $S
\subseteq_f T$, we would have a $\Sigma^0_2$ interpolant between $T$
and $S$ by the compactness theorem. In the next two sections, we show
that this weaker question indeed has a positive answer by showing that
theories in $PSC(k)$ (and $PSC_f$) are equivalent to $\Sigma^0_2$
theories.  This is done in two stages.  In Section
\ref{section:characterizing-Sigma_n-theories}, we give a new semantic
characterization of $\Sigma^0_n$ theories (and hence $\Sigma^0_2$
theories) via a preservation property that we call as
\emph{preservation under $\Sigma^0_n$-closures}. In Section
\ref{section:theories-in-PSCf-and-PSC(k)-are-equivalent-to-Sigma_2-theories},
we show that theories in $PSC_f$ and $PSC(k)$ are preserved under
$\Sigma^0_2$-closures, whence they would be equivalent to $\Sigma^0_2$
theories.

\section{Preservation under $\Sigma^0_n$-closures -- A Semantic Characterization of $\Sigma^0_n$ theories}\label{section:characterizing-Sigma_n-theories}
As stated in the beginning of the report, by $\tau$ we will always
mean a finite vocabulary.

We first introduce some notations. Given a $\tau$-structure $M$, we
denote by $\tau_M$, the vocabulary obtained by expanding $\tau$ with
$|M|$ fresh constants -- one constant per element of $M$. Given a
$\tau$-structure $R$ such that $M \subseteq R$, we denote by $R_M$,
the $\tau_M$ structure whose $\tau$-reduct is $R$ and in which the
constant in $\tau_M \setminus \tau$ corresponding to an element $a$ of
$M$ is interpreted as $a$. In particular therefore, $M_M$ is a
$\tau_M$-structure whose $\tau$-reduct is $M$ and in which every
element of the universe is an interpretation for some constant in
$\tau_M \setminus \tau$. By $S_{(\Sigma, n)}(M)$, we mean the set of
all $\Sigma^0_n$ sentences true in $M$. Likewise, $S_{(\Pi, n)}(M)$
denotes the set of all $\Pi^0_n$ sentences true in $M$.  With the
above notations, one can see that for any structure $M$, the diagram
of $M$, denoted $\text{Diag}(M)$, can be defined as $\text{Diag}(M) =
S_{(\Sigma, 0)}(M_M) = S_{(\Pi, 0)}(M_M)$ and the elementary diagram
of $M$, denoted $\text{El-diag}(M)$, can be defined as
$\text{El-diag}(M) = \bigcup_{n \ge 0} S_{(\Sigma, n)}(M_M) =
\bigcup_{n \ge 0} S_{(\Pi, n)}(M_M)$. By $M \Rrightarrow_n R$, we mean
that every $\Sigma^0_n$ sentence that is true in $M$ is also true in
$R$. Equivalently, every $\Pi^0_n$ sentence that is true in $R$ is
also true in $M$. Using the notation just introduced, $M
\Rrightarrow_n R$ iff $R \models S_{(\Sigma, n)}(M)$ iff $M \models
S_{(\Pi, n)}(R)$. By $M \equiv_n R$, we mean that $M$ and $R$ agree on
all sentences in $\Sigma^0_n$ and $\Pi^0_n$. That is, $M \equiv_n R$
iff $M \Rrightarrow_n R$ and $R \Rrightarrow_n M$.  Finally, $M
\preceq_n R$ denotes that $M \subseteq R$ and for every $\Sigma^0_n$
formula $\varphi(x_1, \ldots, x_r)$ and every $r$-tuple $\bar{a}$ from
$M$, $R \models \varphi(\bar{a})$ iff $M \models
\varphi(\bar{a})$. Using the notation above, $M \preceq_n R$ iff $M
\subseteq R$ and $R_M \equiv_n M_M$.  Using the fact that $\Sigma^0_n$
formulae contain both $\Sigma^9_{n-1}$ and $\Pi^0_{n-1}$ formulae, it
can be shown easily that $\equiv_n$ in the definition of $\preceq_n$
can be replaced with $\Rrightarrow_n$ to get an equivalent definition
of $\preceq_n$. Thus, $M \preceq_n R$ iff $M \subseteq R$ and $R_M
\Rrightarrow_n M_M$. Observe that $M \preceq_0 R$ iff $M \subseteq R$.

The notations $R_M$, $\preceq_n$ and $\Rrightarrow_n$ already appear
in the literature (see~\cite{hodges}) and mean the same as what we
have mentioned above. The notations $S_{(\Sigma, n)}(M)$ and $S_{(\Pi,
  n)}(M)$ have been introduced by us since they will be referred to in
a number of places in the remainder of this report.

It is easy to see $M \preceq_n R$ implies that $M \preceq_{n-1}
R$. Next, $M \equiv R$ iff $M \equiv_n R$ for each $n \in
\mathbb{N}$. Likewise, $M \preceq R$ iff $M \preceq_n R$ for each $n
\in \mathbb{N}$. For $n = 1$, the notion of $\preceq_n$ has a special
name in the literature (see~\cite{chang-keisler}): $M$ is said to be
\emph{existentially closed (e.c.) in} $R$ iff $M \preceq_1 R$. There
are many studies in the literature on this notion. The reader is
referred to~\cite{chang-keisler} for details.

We provide below are some results concerning the $\preceq_n$
relation. We have not encountered these in the literature but these
are simple generalizations of the corresponding results in the
literature concerning e.c. (see~\cite{chang-keisler}, Pg. 192).

In all of the results below, $n \in \mathbb{N}$ i.e. $n$ is a
non-negative integer.

\begin{lemma}\label{lemma:preceq_n-equiv-to-Rrightarrow_{n+1}}
$M \preceq_n R$ iff $M \subseteq R$ and $M_M \Rrightarrow_{n+1} R_M$.
\end{lemma}
\ti{Proof}: The `If' direction is easy to see since $\Sigma^0_{n+1}$
formulas include $\Sigma^0_n$ formulas and $\Pi^0_n$ formulas. For the
`Only if' direction, consider a $\Sigma^0_{n+1}$ $FO(\tau_M)$ sentence
$\varphi = \exists \bar{y} \psi(\bar{y})$ where $\bar{y}$ is an
$n$-tuple and $\psi(\bar{y})$ is a $\Pi^0_n$ formula of
$FO(\tau_M)$. Suppose $M_M \models \varphi$. Then there is an
$n$-tuple $\bar{a} = (a_1, \ldots, a_n)$ from $M_M$ s.t. $M_M \models
\psi(\bar{a})$. Consider the $FO(\tau_M)$ sentence $\xi =
\psi\left[x_1 \mapsto c_1; \ldots; x_n \mapsto c_n \right]$ where
$c_i$ is the constant corresponding to $a_i$ in $\tau_M$. Then $M_M
\models \xi$.  Now since $M \preceq_n R$ and $\xi$ is a $\Pi^0_n$
sentence of $FO(\tau_M)$, we have $R_M \models \xi$. Then $R_M \models
\psi(\bar{a})$, whence $R_M \models \varphi$.\epf

\begin{lemma}\label{lemma:sandwich-result-for-preceq_n}
If $M \preceq_{n+1} R$ and $M \subseteq P \preceq_n R$, then $M
\preceq_{n+1} P$.
\end{lemma}
\ti{Proof}: Suppose $P_M \models \varphi$ where $\varphi$ is a
$\Sigma^0_{n+1}$ sentence of $FO(\tau_M)$. Then since $\varphi$ is
also a $FO(\tau_P)$ sentence, $P_P \models \varphi$. Since $P
\preceq_n R$, we have by Lemma
\ref{lemma:preceq_n-equiv-to-Rrightarrow_{n+1}} that $P_P
\Rrightarrow_{n+1} R_P$ and hence $R_P \models \varphi$. Then since
$\varphi \in FO(\tau_M)$, $R_M \models \varphi$. Finally since $M
\preceq_{n+1} R$, we have $M_M \models \varphi$. \epf\\

The next lemma gives a kind of converse to Lemma
\ref{lemma:sandwich-result-for-preceq_n}. Recall that $|M|$ denotes
the cardinality of the universe of $M$.

\begin{lemma}\label{lemma:the-preceq_n-chain-lemma}
$M \preceq_{n+1} R$ iff there exists $N$ such that $M \preceq N$, 
  $M \preceq_{n+1} R \preceq_n N$ and $|N| = |R|$.
\end{lemma}
\ti{Proof}: The `If' direction is trivial so suppose that $M \preceq_n
R$. Let $T = S_{(\Pi, n)}(R_R) \cup \text{El-diag}(M)$ where for every
element $a$ of $M$, the constant in $\tau_R$ corresponding to $a$ is
\emph{the same} as the constant in $\tau_M$ corresponding to $a$ (and
hence the constants in $\tau_R \setminus \tau_M$ correspond exactly to
the elements in $R$ that are not in $M$).  Any finite subset of
$S_{(\Pi, n)}(R_R)$, resp. $\text{El-diag}(M)$, is satisfiable by
$R_R$, resp. $M$. Hence consider any finite subset $Z$ of $T$ which
has a non-empty intersection with both $S_{(\Pi, n)}(R_R)$ and
$\text{El-diag}(M)$ . Since $S_{(\Pi, n)}(R_R)$ and
$\text{El-diag}(M)$ are each closed under finite conjunctions, we can
consider $Z = \{\xi, \psi\}$ where $\xi \in S_{(\Pi, n)}(R_R)$ and
$\psi \in \text{El-diag}(M)$.  Let $c_1, \ldots, c_r$ be the constants
of $\tau_R \setminus \tau_M$ appearing in $\xi$ and consider $\phi$
given as \linebreak $\phi = \exists x_1 \ldots \exists x_r \xi
\left[c_1 \mapsto x_1; \ldots; c_r \mapsto x_r \right]$. Observe that
$\phi$ is a $\Sigma^0_{n+1}$ sentence of $R_M$ and that $R_M \models
\phi$. Then since $M \preceq_{n+1} R$, we have that $M_M \models
\phi$. Let $a_1, \ldots, a_r$ be the witnesses for the variables $x_1,
\ldots, x_r$ in $\phi$ mentioned above. Then interpreting the
constants $c_1, \ldots, c_r$ as $a_1, \ldots, a_r$, one can check that
$(M_M, a_1, \ldots, a_r) \models Z$. Since $Z$ was arbitrary, by the
compactness theorem, $T$ is satisfiable by a $\tau_R$-structure
$\mc{N}$ whose universe contains the universe of $R$. Let $N$ be the
$\tau$-reduct of $\mc{N}$. Since $\mc{N} \models \text{El-diag}(M)$,
we have $M \preceq N$. Since $\mc{N} \models S_{(\Pi, n)}(R_R)$, we
have $R \preceq_n N$. Finally as for the size of $N$, we have two
cases. If $M$ is finite, then since $M \preceq N$, we have $M = R = N$
so that trivially $|N| = |R|$. Else if $M$ is infinite, then by the
L\"owenheim-Skolem theorem, $\mc{N}$ can be chosen such that $|\mc{N}|
= |R|$, whence $|N| = |R|$. In either case, therefore $|N| =
|R|$.\epf

\begin{corollary}
For $n \ge 1$, $M \preceq_{n+1} R$ iff there exists $N_1, \ldots,
N_{n+1}$ such that (i) $M \preceq N_1$ and $R \preceq N_2$ (ii) $N_1
\preceq N_3 \preceq \ldots \preceq N_i$ where $i = 2 \cdot
\lfloor\frac{n}{2}\rfloor + 1$ and $N_2 \preceq N_4 \preceq \ldots
\preceq N_j$ where $j = 2 \cdot \lfloor\frac{(n+1)}{2}\rfloor$ (iii)
$M \preceq_{n+1} R \preceq_n N_1 \preceq_{n-1} N_2 \preceq_{n-2}
\ldots \preceq_1 N_n \preceq_0 N_{n+1}$ (iv) $|N_{n+1}| = |N_n| =
\ldots = |N_1| = |R|$.
\end{corollary}

As mentioned at the start of this section, the $\preceq_1$ relation
has a special name in the literature: $M$ is said to be
\emph{existentially closed (e.c.)} in $N$ iff $M \preceq_1 R$. There
is a natural generalization of this notion to general $n$ as follows:
We say that $M$ is \emph{$\Sigma^0_n$-closed in} $R$ iff $M \preceq_n
R$. This notion generalizes the notion of existential closure since
$M$ is e.c. in $R$ iff $M$ is $\Sigma^0_1$-closed in $R$. We call $R$
as a \emph{$\Sigma^0_n$-closure of} $M$ if $M$ is $\Sigma^0_n$-closed
in $R$.  We can now talk about the following preservation property.

\begin{defn}[Preservation under $\Sigma^0_n$-closures]
A class $S$ of structures is said to be \emph{preserved under
  $\Sigma^0_n$-closures} if for every structure $M \in S$, if $R$ is a
$\Sigma^0_n$-closure of $M$, then $R \in S$. If $S$ is defined by a FO
theory $T$, then we say $T$ is \emph{preserved under
  $\Sigma^0_n$-closures}.
\end{defn}

For the case of $n = 0$, we know that $R$ is a $\Sigma^0_0$-closure of
$M$ iff $M \subseteq R$. Then preservation under $\Sigma^0_0$-closures
is the same as preservation under extensions. Indeed, there is a
syntactic characterization of elementary classes of structures that
are preserved under extensions -- this is given by the {\lt} theorem
which in extensional form is stated as: A FO theory is preserved under
extensions iff it is equivalent to a theory of existential sentences.

The aim of the remainder of this section is to generalize the
extensional form of the {\lt} theorem just mentioned, by providing a
syntactic characterization of preservation under $\Sigma^0_n$-closures
for each $n \in \mathbb{N}$, which would give us exactly the class of
theories of existential sentences when $n = 0$. Our proof is a
generalization of the proof of the {\lt} theorem. The proof of the
{\lt} theorem uses the following key theorem.

\begin{theorem}[Existential Amalgamation Theorem (EAT), ref.~\cite{hodges}]
Let $N, P$ be $\tau$-structures. Then $N \Rrightarrow_1 P$ iff there
exists an elementary extension $R$ of $P$ and an isomorphic copy $M$
of $N$ such that $M \subseteq R$.
\end{theorem}

Towards our proof, we will first prove the following generalization of
the EAT. Observe that the statement of EAT is exactly the statement of
the lemma below for $n = 0$.

\begin{theorem}[Generalization of EAT]\label{theorem:generalizing-JEL-for-Sigma_n}
Let $N, P$ be $\tau$-structures and $n \in \mathbb{N}$. Then $N
\Rrightarrow_{n+1} P$ iff there exists an elementary extension $R$ of
$P$ and an isomorphic copy $M$ of $N$ such that $M \preceq_n R$.
\end{theorem}
\ti{Proof}: Choose $\tau_P$ and $\tau_N$ such that $\tau_P \cap \tau_N
= \tau$.  Consider the $\tau_N$-structure $N_N$. Consider $Z =
\text{El-diag}(P) \cup S_{(\Pi, n)}(N_N)$, which is a theory in
$\tau_P \cup \tau_N$. Suppose $Z$ is unsatisfiable. Since $S_{(\Pi,
  n)}(N_N)$ is closed under finite conjunctions, by compactness
theorem, there is a sentence $\psi \in S_{(\Pi, n)}(N_N)$ such that
$\text{El-diag}(P) \cup \{\psi\}$ is unsatisfiable.  Now let $c_1,
\ldots, c_r$ be the constants of $\tau_N \setminus \tau$ appearing in
$\psi$ and let $\varphi = \exists x_1 \ldots \exists x_r \psi\left[c_1
  \mapsto x_1; \ldots; c_r \mapsto x_r \right]$. It is easy to see
that $\varphi$ is a $\Sigma^0_{n+1}$ sentence of $FO(\tau)$ and also
that $N \models \varphi$. Since $N \Rrightarrow_{n+1} P$, we have $P
\models \varphi$. Now $\text{El-diag}(P) \rightarrow \neg \psi$. Since
the constants $c_1, \ldots, c_r$ do not appear in $\tau_P$, we have by
$\forall$-introduction that $\text{El-diag}(P) \rightarrow \forall x_1
\ldots \forall x_r \neg \psi\left[c_1 \mapsto x_1; \ldots; c_r \mapsto
  x_r \right] = \neg \varphi$. Then $P \models \neg \varphi$. This
contradicts the earlier inference that $P \models \varphi$.

Then $Z$ is satisfiable by a $(\tau_P \cup \tau_N)$-structure, say
$\mathbf{R} = (R, a_1, \ldots, a_\alpha, b_1, \ldots, b_\beta)$ where
(i) $R$ is a $\tau$-structure (ii) $a_1, \ldots, a_\alpha$ are the
interpretations of the constants in $\tau_P \setminus \tau$ and
$\{a_1, \ldots, a_\alpha\}$ is exactly the universe of $P$ and finally
(iii) $b_1, \ldots, b_\beta$ are the interpretations of the constants
in $\tau_N \setminus \tau$ and $\beta = |N|$. Then $P \preceq
R$. Since $(R, b_1, \ldots, b_\beta) \models Z$, and hence $S_{(\Pi,
  n)}(N_N)$, and since $\text{Diag}(N_N) \subseteq S_{(\Pi, n)}(N_N)$,
it follows that the substructure $M$ of $R$ formed by $b_1, \ldots,
b_\beta$ is isomorphic to $N$ and that the universe of $M$ is exactly
$\{b_1, \ldots, b_\beta\}$. Since $M$ is a $\tau$-structure which is
isomorphic to $N$, we can treat $\tau_M$ and $\tau_N$ as
identical. Then $N_N \cong M_M$ and the structure $(R, b_1, \ldots,
b_\beta)$ can be treated as $R_M$.

We will now show that $M \preceq_n R$ to complete the proof.

Let $\phi$ be a $\Sigma^0_n$ sentence of $FO(\tau_M)$ true in
$R_M$. Since $\mathbf{R} \models S_{(\Pi, n)}(N_N)$ and since $\tau_M
= \tau_N$ , we have $R_M \models S_{(\Pi, n)}(N_N)$. Then $N_N \models
\phi$, whence $M_M \models \phi$ since $N_N \cong M_M$. \epf\\

Putting $n = 1$ in the statement of the above lemma, we get the
following.

\begin{corollary}\label{corollary:generalizing-JEL-for-Sigma_2}
Let $N, P$ be $\tau$-structures. Then $N \Rrightarrow_2 P$ iff there
exists an elementary extension $R$ of $P$ and an isomorphic copy $M$
of $N$ such that $M$ is e.c. in $R$.
\end{corollary}

A slight digression before proceeding ahead: By a little modification
of the proof of Theorem \ref{theorem:generalizing-JEL-for-Sigma_n}, in
particular, by considering $\bigcup_{n \ge 0} S_{(\Pi, n)}(N_N)$
instead of $S_{(\Pi, n)}(N_N)$, we can show the following.

\begin{theorem}\label{theorem:generalizing-JEL-for-all-of-FO}
Let $N, P$ be $\tau$-structures. Then $N \equiv P$ iff there exists an
elementary extension $R$ of $P$ and an isomorphic copy $M$ of $N$ such
that $M \preceq R$.
\end{theorem}

This shows that though two elementarily equivalent structures may not
be related by an elementary substructure relation, it is always
possible to elementarily extend one of them such that the extension
contains an isomorphic copy of the other structure as an elementary
substructure.\\

Having proved Theorem \ref{theorem:generalizing-JEL-for-Sigma_n}, wee
are now ready to give a syntactic characterization of preservation
under $\Sigma^0_n$-closures.

\begin{theorem}\label{theorem:characterizing-Sigma_{n+1}-theories-using-pres-under-Sigma_n-closures}
For each $n \ge 0$, a theory $T$ is preserved under
$\Sigma^0_n$-closures iff $T$ is equivalent to a theory of
$\Sigma^0_{n+1}$ sentences.
\end{theorem}
\ti{Proof}: 

\underline{If}: Let $Z$ be a theory of $\Sigma^0_{n+1}$
sentences. Suppose $M \models Z$ and $R$ is a $\Sigma^0_n$-closure of
$M$. Consider any sentence $\phi \in Z$. Then $\phi$ is of the form
$\exists x_1 \ldots \exists x_n \xi(x_1, \ldots, x_n)$ where $\xi$ is
a $\Pi^0_n$ formula of $FO(\tau)$. Since $M \models \phi$, there
exists an $n$-tuple $\bar{a}$ from $M$ such that $M \models
\xi(\bar{a})$. Then since $M$ is $\Sigma^0_n$-closed in $R$, we have
$R \models \xi(\bar{a})$ (by definition). Then $R \models \phi$. Since
$\phi$ was an arbitrary sentence of $Z$, we have $R \models Z$.

\underline{Only If}: Let $\Gamma$ be the set of all $\Sigma^0_{n+1}$
consequences of $T$. Clearly $T \rightarrow \Gamma$. In the converse
direction, suppose $P \models \Gamma$. We will show that $P \models T$
to complete the proof.

Consider the $FO(\tau)$ theory $Z = T \cup S_{(\Pi, n+1)}(P)$ and
suppose that $Z$ is unsatisfiable. Since $S_{(\Pi, n+1)}(P)$ is closed
under finite conjunctions, by compactness theorem, there is $\phi \in
S_{(\Pi, n+1)}(P)$ such that $T \cup \{\phi\}$ is unsatisfiable. Then
$T \rightarrow \neg \phi$. If $\varphi = \neg \phi$, then $\varphi$ is
equivalent to a $\Sigma^0_{n+1}$ sentence. Then $\varphi \in \Gamma$
and hence $P \models \varphi$. Then $P \not\models \phi$ -- a
contradiction.

Then $Z$ is satisfiable in a structure say $N$. Hence, $N \models T$
and $N \Rrightarrow_{n+1} P$. By Theorem
\ref{theorem:generalizing-JEL-for-Sigma_n}, there exists an elementary
extension $R$ of $P$ and an isomorphic copy $M$ of $N$ such that $M
\preceq_n R$. Then $M$ is $\Sigma^0_n$-closed in $R$. Since $N \models
T$ and $M \cong N$, we have $M \models T$. Since $T$ is preserved
under $\Sigma^0_n$-closures and since $R$ is a $\Sigma^0_n$-closure of
$M$, we have $R \models T$. Finally since $P \preceq R$, we have $P
\models T$. \epf\\

Putting $n = 0 $ in the above result, we get the extensional form of
the {\lt} theorem. Theorem
\ref{theorem:characterizing-Sigma_{n+1}-theories-using-pres-under-Sigma_n-closures}
therefore gives us a different generalization of the extensional form
of the {\lt} theorem than the one given by Theorem
\ref{theorem:generalizations}.\\
 
\underline{\textbf{A Comparison with Literature}}:

While we have not encountered the above characterization of
$\Sigma^0_n$ theories in the literature, we remark that the motivation
came from our trying to prove the following theorem from~\cite{hodges}
where the theorem is given as an exercise problem.

\begin{theorem}[ref.~\cite{hodges}, Chp. 5, Sect. 5.4]
A theory $T$ is equivalent to a $\Sigma^0_2$ theory iff for all
structures $M$, $N$ and $R$, if $M$ and $N$ are models of $T$ such
that $M \preceq N$ and $M \subseteq R \subseteq N$, then $R$ is also a
model of $T$.
\end{theorem}

Now suppose the semantic condition, call it $\mc{C}$, mentioned in the
theorem above is true for a theory $T$. Suppose $M \models T$ and $M$
is $\Sigma^0_1$-closed in $R$ (i.e. $M$ is e.c. in $R$). Then from
Lemma \ref{lemma:the-preceq_n-chain-lemma} and the fact that the
$\preceq_0$ relation is the same as the $\subseteq$ relation, there
exists $N$ such that $M \preceq N$ and $M \subseteq R \subseteq
N$. Then $\mc{C}$ tells us that $R \models T$. In other words, $T$ is
preserved under $\Sigma^0_1$-closures. Conversely, suppose $T$ is
preserved under $\Sigma^0_1$-closures and suppose the precondition of
$\mc{C}$ is true for structures $M$, $N$ and $R$. Then $M$ is e.c. in
$R$ by Lemma \ref{lemma:sandwich-result-for-preceq_n}. Since $M
\models T$ and $T$ is preserved under $\Sigma^0_1$-closures, we have
$R \models T$. To sum up, the semantic condition $\mc{C}$ can be more
succinctly reworded as preservation under
$\Sigma^0_1$-closures. Further, for each $n \ge 0$, this rewording
lends itself to a generalization that captures exactly the class of
all theories of $\Sigma^0_n$ sentences!

\section{Theories in $PSC_f$ and $PSC(k)$ are equivalent to $\Sigma^0_2$ theories}\label{section:theories-in-PSCf-and-PSC(k)-are-equivalent-to-Sigma_2-theories}

In this section, we will prove that for each theory in $PSC_f$ and
$PSC(k)$ for each $k \in \mathbb{N}$, there exists an equivalent
theory consisting of only $\Sigma^0_2$ sentences. This result
therefore makes partial progress on the problem posed in Section
\ref{section:generalizations}, of getting a syntactic characterization
of theories in $PSC_f$ and $PSC(k)$.

We repeat below some of the results from the literature that we had
recalled in Section \ref{section:proof-of-conjecture}, for the sake of
convenience of reading and quick reference.
\vspace{7pt}

\textbf{Theorem \ref{theorem:existence-of-saturated-elementary-extensions}~(ref.~\cite{chang-keisler}, repeated from Section \ref{section:proof-of-conjecture}).} \textit{Let $\tau$ be a finite vocabulary, $\lambda$ be an infinite cardinal
and $M$ be a $\tau$-structure such that $\omega \leq |M| \leq
2^{\lambda}$ ($|M|$ denotes the power of $M$). Then there is a
$\lambda$-saturated elementary extension of $M$ of power $2^\lambda$.}
\vspace{7pt}

\textbf{Proposition \ref{proposition:extending-saturations-to-expansions}~(ref.~\cite{chang-keisler}, repeated from Section \ref{section:proof-of-conjecture}).}
\textit{Given an infinite cardinal $\lambda$ and a $\lambda$-saturated
$\tau$-structure $M$, for every $k$-tuple $(a_1, \ldots, a_k)$ of
elements from $M$ where $k \in \mathbb{N}$, the $\tau_k$ structure
$(M, a_1, \ldots, a_k)$ is also $\lambda$-saturated.}
\vspace{7pt}

\textbf{Proposition \ref{proposotion:finite-structures-are-saturated}~(ref.~\cite{chang-keisler}, repeated from Section \ref{section:proof-of-conjecture}).}
\textit{A $\tau$-structure is finite iff it is $\lambda$-saturated for
all cardinals $\lambda$.}
\vspace{7pt}

\textbf{Theorem \ref{theorem:embedding-conditions}~(ref.~\cite{chang-keisler}, repeated from Section \ref{section:proof-of-conjecture}).}
\textit{Given $\tau$-structures $M$ and $N$ and a cardinal $\lambda$, suppose
that (i) $M$ is $\lambda$-saturated, (ii) $\lambda \ge |N|$, and (iii)
$N \Rrightarrow_1 M$. Then $N$ is isomorphically embeddable in $M$.}
\vspace{7pt}

The following result is a simple extension of a result from
~\cite{chang-keisler}.
\begin{proposition}\label{proposition:saturated-implies-realizability-of-any-set-of-formulas}
Let $M$ be an infinite structure that is $\alpha$-saturated for some
$\alpha \ge \omega$. Then for each subset $Y$ of the universe of $M$,
of size $< \alpha$, each set of formulas $\Gamma(x_1, \ldots, x_k)$ of
$\tau_Y$ that is consistent with $\text{Th}(M_Y)$ is realized in
$M_Y$.
\end{proposition}

Before proceeding with the technical details, we first give the
outline of our proof. From the characterization of $\Sigma^0_n$
theories given by Theorem
\ref{theorem:characterizing-Sigma_{n+1}-theories-using-pres-under-Sigma_n-closures}
in the previous section, we know that a theory $T$ is equivalent to a
theory of $\Sigma^0_2$ sentences iff $T$ is preserved under
$\Sigma^0_1$-closures. Therefore to show that a theory $T$ in $PSC_f$
or $PSC(k)$ has an equivalent theory consisting of only $\Sigma^0_2$
sentences, we show that theories in $PSC_f$ and $PSC(k)$ are preserved
under $\Sigma^0_1$-closures. This is achieved as follows: To show that
$(M_1 \models T$ and $M_1 \preceq_1 R_1) \rightarrow R_1 \models T$
for structures $M_1$ and $R_1$, we show that there exist structures
$M$ and $R$ such that $M$ is $\alpha$-saturated for some $\alpha \ge
\omega$, $M \equiv M_1$, $R \equiv R_1$ and $M \preceq_1 R$. This is
proved in Lemma
\ref{lemma:lifting-the-e.c.-relation-to-an-elem-equivalent-saturated-e.c.-relation}
below. It then suffices to prove that $T$ is preserved under
$\Sigma^0_1$-closures of $\alpha$-saturated models for each $\alpha
\ge \omega$. This is shown in Lemma
\ref{lemma:closure-of-PSC(k)-and-PSC_f-under-e.c.-for-saturated-ec.-relation}. And
that completes the proof.\\

We now give all the technical details.  Recall that $M$ is said to be
existentially closed (e.c.) in $R$ if $M \preceq_1 R$.

\begin{lemma}\label{lemma:lifting-the-e.c.-relation-to-an-elem-equivalent-saturated-e.c.-relation}
Let $M_1$ and $R_1$ be infinite $\tau$-structures such that $M_1$ is
e.c. in $R_1$. Let $\alpha = |M_1|$ and $\beta = \text{max}(|R_1|,
2^\alpha)$. Then there exist $\tau$-structures $M$ and $R$ such that
(i) $|M| = 2^{\alpha}$ and $|R| = \beta$ (ii) $M$ is
$\alpha$-saturated and $M$ is e.c. in $R$ (iv) $M_1 \preceq M$ and
$R_1$ is elementarily embeddable in $R$.
\end{lemma}
\ti{Proof}: By Theorem
\ref{theorem:existence-of-saturated-elementary-extensions}, there
exists a $\alpha$-saturated elementary extension $M$ of $M_1$ of power
$2^\alpha$.  Consider $T = S_{(\Pi, 1)}(M_M) \cup \text{El-diag}(R_1)$
where $\tau_M \cap \tau_{R_1} = \tau$. Consider any finite subset of
$Z$ of $T$. If $Z \subseteq S_{(\Pi, 1)}(M_M)$ or $Z \subseteq
\text{El-diag}(R_1)$, then $Z$ is clearly satisfiable. Else $Z = V
\cup W$ where $\emptyset \neq V \subseteq S_{(\Pi, 1)}(M_M)$ and
$\emptyset \neq W \subseteq \text{El-diag}(R_1)$. Since each of
$S_{(\Pi, 1)}(M_M)$ and $\text{El-diag}(R_1)$ is closed under finite
conjunctions, $V = \{\xi\}$ and $W = \{\psi\}$. Let $\chi(x_1, \ldots,
x_r) = \xi \left[c_1 \mapsto x_1; \ldots; c_r \mapsto x_r\right]$
where $c_1, \ldots, c_k$ are the constants of $\tau_M \setminus \tau$
appearing in $\xi$. Let $\phi = \exists x_1 \ldots \exists x_r
\chi(x_1, \ldots, x_r)$. Since $\chi$ is a $\Pi^0_1$ formula, $\phi$
is a $\Sigma^0_2$ sentence of $FO(\tau)$. It is clear that $M \models
\phi$. Since $M_1 \preceq M$, we have $M_1 \models \phi$. Then $M_1
\models \chi(\bar{a})$ for some $r$-tuple $\bar{a}$ from $M_1$. Since
$M_1$ is e.c. in $R_1$, we have $R_1 \models \chi(\bar{a})$. Then one
can see that if $P = R_1$, then $(P_P, \bar{a}) \models Z$. Since
every finite subset of $T$ is satisfiable, by the Compactness theorem,
$T$ is satisfiable. By the L\"owenheim-Skolem theorem, there exists a
$\tau$-structure $R$ of power $\beta$ such that (i) $M \subseteq R$
and $M$ is e.c. in $R$ (ii) $R_1$ is elementarily embeddable in
$R$.\epf

\begin{lemma}\label{lemma:closure-of-PSC(k)-and-PSC_f-under-e.c.-for-saturated-ec.-relation}
Let $S$ be a class of structures in $\PSCf$ or ${\PSC}(k)$ for some $k
\in \mathbb{N}$, that is closed under elementary extensions and
isomorphisms. Let $M$ and $R$ be infinite $\tau$-structures such that
(i) $M \in S$ (ii) $M$ is $\alpha$-saturated for some $\alpha \ge
\omega$ and (iii) $M$ is e.c. in $R$. Then $R \in S$.
\end{lemma}
\ti{Proof}: We give the proof for $S \in \PSCf$. The proof for $S \in
   {\PSC}(k)$ is similar.

Since $M$ is e.c. in $R$, by Lemma
\ref{lemma:the-preceq_n-chain-lemma}, there exists a $\tau$-structure
$N_1$ such that $M \preceq N_1$ and $M \preceq_1 R \subseteq N_1$. By
Theorem \ref{theorem:existence-of-saturated-elementary-extensions},
there exists a $\beta$-saturated elementary extension $N$ of $N_1$, of
power $2^\beta$, for some $\beta \ge |N_1|$. Since $M \preceq N_1$ and
$N_1 \preceq N$, we have $M \preceq N$. Since $M \in S$ and $S$ is
closed under elementary extensions, we have $N \in S$. Now $S$ being
in $\PSCf$, there exists a finite core $C$ of $N$ of size say $k$. Let
$\bar{a}$ be any $k$-tuple formed from $C$. Let $\Gamma(x_1, \ldots,
x_k)$ be the FO-type of $\bar{a}$ in $N$. Now since $M \preceq N$,
$\text{Th}(M) = \text{Th}(N)$. Then $\Gamma(x_1, \ldots, x_n)$ is
consistent with $\text{Th}(M)$. Since $M$ is $\alpha$-saturated, from
Proposition
\ref{proposition:saturated-implies-realizability-of-any-set-of-formulas},
$\Gamma(x_1, \ldots, x_k)$ is realized in $M$ by a tuple say
$\bar{b}$. Then since $M \preceq N$, $N \models \Gamma(\bar{b})$. Then
for every universal formula $\varphi(x_1, \ldots, x_k) \in \Gamma(x_1,
\ldots, x_k)$, we have $R \models \varphi(\bar{b})$ since universal
formulae are preserved under substructures by the {\lt} theorem. Which
means that every universal sentence true in $(N, \bar{a})$ is also
true in $(R, \bar{b})$. In other words, $(R, \bar{b}) \Rrightarrow_1
(N, \bar{a})$. Now since $N$ is $\beta$-saturated, $(N, \bar{a})$ is
also $\beta$-saturated by Proposition
\ref{proposition:extending-saturations-to-expansions}. Further since
$\beta \ge |N_1|$ and $R \subseteq N_1$, we have $\beta \ge |R|$,
whence $\beta \ge |(R, \bar{b})|$. Then by Theorem
\ref{theorem:embedding-conditions}, $(R, \bar{b})$ is isomorphically
embeddable in $(N, \bar{a})$ via an isomorphic embedding $f$. Then the
image of $(R, \bar{b})$ under $f$ is a structure $(R_1, \bar{a})$ such
that $R_1$ is a $\tau$-structure, $R_1 \subseteq N$ and $R_1$ contains
$C$. Since $C$ is a core of $N$, $R_1 \in S$ by definition. Finally,
since $R_1$ is isomorphic to $R$ and $S$ is closed under isomorphisms,
we have $R \in S$. \epf

\begin{lemma}\label{lemma:closure-of-PSCf-and-PSC(k)-under-e.c.}
Let $S$ be a class of structures in $\PSCf$ or ${\PSC}(k)$ for some $k
\in \mathbb{N}$, that is closed under elementary extensions,
elementary substructures and isomorphisms. Then $S$ is preserved under
$\Sigma^0_1$-closures.
\end{lemma}
\ti{Proof}: 
Suppose $M_1 \in S$ and $M_1$ is e.c. in $R_1$. We will show that
$R_1 \in S$.

If $M_1$ is finite, then suppose it has $n$ elements, say
$a_1, \ldots, a_n$. Let $\phi(x_1, \ldots, x_n)$ $ = \forall
y \bigvee_{i = 1}^{i = n} (y = x_i)$. Now observe that
$M_1 \models \phi(a_1, \ldots, a_n)$. Since $M_1$ is e.c. in $R_1$ and
since $\phi$ is a $\Pi^0_1$ formula, $R_1 \models \phi(a_1, \ldots,
a_n)$. Then $M_1 = R_1$, whence $R_1 \in S$.

Else $M_1$ and $R_1$ are both infinite.  Then by Lemma
\ref{lemma:lifting-the-e.c.-relation-to-an-elem-equivalent-saturated-e.c.-relation}
there exist structures $M$ and $R$ such that (i) $M$ is
$\alpha$-saturated for some $\alpha \ge \omega$ (ii) $M$ is e.c. in
$R$ (iii) $M_1 \preceq M$ and $R_1$ is elementarily embeddable in
$R$. Since $S$ is closed under elementary extensions and $M_1 \in S$,
we have $M \in S$. Then invoking Lemma
\ref{lemma:closure-of-PSC(k)-and-PSC_f-under-e.c.-for-saturated-ec.-relation}, 
we get $R \in S$.  Finally, since $S$ is closed under elementary
substructures and isomorphisms, and hence elementary embeddings,
$R_1 \in S$. \epf.

\begin{corollary}\label{corollary:theories-in-PSCf-and-PSC(k)-are-closed-under-e.c.}
Let $T$ be a theory in $PSC_f$ or $PSC(k)$ for some $k \in
\mathbb{N}$. Then $T$ is preserved under $\Sigma^0_1$-closures.
\end{corollary}
\ti{Proof}:
Since $T$ is a theory, it is closed under elementary extensions,
elementary substructures and isomorphisms. Invoking Lemma
\ref{lemma:closure-of-PSCf-and-PSC(k)-under-e.c.}, we are done.\epf 

\begin{theorem}\label{theorem:Sigma_2-theories-subsume-PSCf-and-PSC(k)-theories}
Let $T$ be a theory in $PSC_f$ or $PSC(k)$ for some $k \in
\mathbb{N}$. Then $T$ is equivalent to a theory of $\Sigma^0_2$
sentences.
\end{theorem}
\ti{Proof}:
By
Corollary \ref{corollary:theories-in-PSCf-and-PSC(k)-are-closed-under-e.c.},
$T$ is preserved under $\Sigma^0_1$-closures. Then by
Theorem \ref{theorem:characterizing-Sigma_{n+1}-theories-using-pres-under-Sigma_n-closures},
$T$ is equivalent to a theory of $\Sigma^0_2$ sentences.\epf\\

\underline{\textbf{Some other results:}}\\

Using the ideas in the proofs above and in the previous subsection, we
can prove the following result which is closely related to
Lemma \ref{lemma:the-preceq_n-chain-lemma}, but which involves
saturated structures.

\begin{lemma}[Sandwich by saturated structures]\label{lemma:sandwich-by-saturated-structures}
Let $M_1$ and $R_1$ be $\tau$-structures such that $M_1 \preceq_n
R_1$.  Then there exist $\tau$-structures $M, N$ and $R$ such that (i)
$M$ is $\alpha$-saturated and $N$ is $\beta$-saturated for some
$\alpha \ge \omega$ and some $\beta \ge \alpha$ (ii) $M \preceq N$,
$M \preceq_n R \preceq_{n-1} N$ and (iii) $M_1 \preceq M$ and $R_1$ is
elementarily embeddable in $R$. If $M_1$ is infinite, then $\alpha$,
$\beta$, $M$, $N$ and $R$ can be chosen such that $\alpha = |M_1|$,
$\beta = \text{max}(|R_1|, 2^\alpha)$, $|M| = 2^{\alpha}$, $|R|
= \beta$ and $|N| = 2^\beta$.
\end{lemma} 
\ti{Proof}: 
If $M_1$ is finite, then in a manner similar to that in the proof of
Lemma
\ref{lemma:closure-of-PSCf-and-PSC(k)-under-e.c.}, we can show that
$M_1 = R_1$. Since by
Proposition \ref{proposotion:finite-structures-are-saturated} every
finite structure is $\alpha$-saturated for all cardinals $\alpha$, we
can choose $M = N = R = M_1$ and see that they are indeed as desired.

Else $M_1$ is infinite. By exactly the same kind of arguments as
presented in the proof of Lemma
\ref{lemma:lifting-the-e.c.-relation-to-an-elem-equivalent-saturated-e.c.-relation}
(in fact just do the following replacements in the latter proof:
$S_{(\Pi, 1)} \mapsto S_{(\Pi, n)};~ \Pi^0_1 \mapsto \Pi^0_n;~
\Sigma^0_2 \mapsto \Sigma^0_{n+1};~ \text{`is e.c. in'} \mapsto
\preceq_n$), we can show that there exist $M$, $R$, $\alpha$ and
$\beta$ such that (i) $M$ is $\alpha$-saturated and $M \preceq_n R$
(ii) $\alpha = |M_1|$, $\beta = \text{max}(|R_1|, 2^\alpha)$, $|M| =
2^{\alpha}$, $|R| = \beta$ and (iii) $M_1 \preceq M$ and $R_1$ is
elementarily embeddable in $R$.  To get $N$ as desired, we see that
since $M \preceq_n R$, by Lemma \ref{lemma:the-preceq_n-chain-lemma},
there exists $N_1$ such that $M \preceq N_1$, $M \preceq_n R
\preceq_{n-1} N_1$ and $|N_1| = |R| = \beta$. Then by Theorem
\ref{theorem:existence-of-saturated-elementary-extensions}, there
exists a $\beta$-saturated elementary extension $N$ of $N_1$ of power
$2^\beta$. Since $R \preceq_n N_1$, we have $R \preceq_n N$. Finally,
since $M \preceq N_1$, we have $M \preceq N$.\epf

\subsection{$\Sigma^0_2$ theories are more general than theories in $PSC_f$ and $PSC(k)$}

While theories in $PSC_f$ and $PSC(k)$ are equivalent to $\Sigma^0_2$
theories, the vice-versa unfortunately is not true. In fact, the
following lemma reveals a dark fact -- even theories of
$\exists \forall^*$ sentences can go beyond $PSC_f$!

\begin{lemma}\label{lemma:a-theory-of-Sigma^0_2-sentences-not-in-PSCf}
There is a theory of $\Sigma^0_2$ sentences in which each sentence has
exactly one existential variable and which is not in $PSC_f$.
\end{lemma}
\ti{Proof}: For $n \ge 1$, let $\varphi_{n}(x)$ be a formula which asserts
that $x$ is not a part of a cycle of length $n$. Explicitly stated,
$\varphi_1(x) = \neg E(x, x)$ and for $n \ge 1$, $\varphi_{n+1}(x)
= \neg \exists z_1 \ldots \exists z_n
\big((\bigwedge_{1 \leq i < j \leq n} z_i \neq z_j) \wedge
(\bigwedge_{i = 1}^{i = n} (x \neq z_i)) \wedge E(x, z_1) \wedge E
(z_n, x) \wedge$ \linebreak $ \bigwedge_{i = 1}^{i = n -1} E(z_i,
z_{i+1})\big)$.

Now consider $\phi_n(x) = \bigwedge_{i = 1}^{i = n} \varphi_i(x)$
which asserts that $\phi_n$ is not a part of any cycle of length $\leq
n$.  Observe that $\phi_n(x) \rightarrow \phi_m(x)$ if $m \leq n$.

Finally consider $\psi_n = \exists x \phi_n(x)$ which asserts that
$\phi_n(x)$ is realized in any model. Let $T = \{ \psi_n \mid n \ge
1\}$. Then $T$ is a theory of $\Sigma^0_2$ sentences in which each
$\Sigma^0_2$ sentence has only one existential variable. We will show
below that $T \notin PSC_f$.

Consider a infinite graph $G$ given by $\coprod C_i$ where $C_i$ is a
cycle of length $i$ and $\coprod$ denotes disjoint union. Any vertex
of $C_i$ satisfies $\phi_j(x)$ for $j < i$, since it is not a part of
any cycle of length $< i$. Then $G \models T$. Now consider any finite
set $S$ of vertices of $G$. Let $k$ be the highest index such that
some vertex in $S$ is in the cycle $C_k$. Then consider the subgraph
$G_1$ of $G$ induced by the vertices of all the cycles in $G$ of
length $\leq k$. Then no vertex of $G$ satisfies $\phi_l(x)$ for $l
\ge k$. Then $G_1 \not\models T$ whence $S$ cannot be a core of
$G$. Since $S$ was an arbitrary finite subset of $G$, we conclude that
$G$ has no finite core. Then $T \notin PSC_f$ (and hence $\notin
PSC(k)$ for any $k \in \mathbb{N}$).\epf\\

This shows that allowing an infinite number of sentences in a
$\Sigma^0_2$ theory to use existential variables can afford power to
the theory to have models that do not have any finite cores. Ofcourse
if the number of $\Sigma^0_2$ sentences using existential variables is
finite, then these sentences can be ``clubbed'' together to get a
single equivalent $\Sigma^0_2$ sentence. Then the original
$\Sigma^0_2$ theory would be equivalent to a $\Sigma^0_2$ theory which
contains only one $\Sigma^0_2$ sentence with the rest of the sentences
being only universal sentences. Such theories are easily seen to be in
$PSC_f$, in fact in $PSC(k)$ where $k$ is the number of variables in
the lone $\Sigma^0_2$ sentence of the theory. For the converse
direction, presently it is unclear if theories in $PSC_f$ and $PSC(k)$
are equivalent to $\Sigma^0_2$ theories in which there is only one
$\Sigma^0_2$ sentence and the rest of the sentences are universal.
However the following question, if resolved positively, would show
that the converse direction is also true: Given a $\Sigma^0_2$ theory
which is \emph{known to be in $PSC_f$, resp. $PSC(k)$}, is it the case
that it is equivalent to a $\Sigma^0_2$ theory containing only one
$\Sigma^0_2$ sentence, resp. only one $\Sigma^0_2$ sentence with $k$
existential variables, and the rest of the sentences are all
universal? If so, then since Theorem
\ref{theorem:Sigma_2-theories-subsume-PSCf-and-PSC(k)-theories} tells
us that theories in $PSC_f$ and $PSC(k)$ are certainly equivalent to
$\Sigma^0_2$ theories (which would also therefore be in $PSC_f$ and
$PSC(k)$ resp.), the special kind of $\Sigma^0_2$ theories mentioned
in the question just stated would give us the desired
characterizations. However, we do not have an answer to this question.

\section{Conclusion and Future Work}\label{section:conclusion}
In this paper, we presented preservation theorems that characterize
the $\exists^k \forall^*$ and $\forall^k \exists^*$ prefix classes of
FO. These theorems can be viewed as generalizations of the
substructual and extensional versions of the {\lt} theorem. Our
results contrast with earlier characterizations of $\Sigma^0_2$ and
$\Pi^0_2$, such as those using ascending chains, descending chains and
$1$-sandwiches, which do not yield the {\lt} theorem as a special
case.  A few open questions remain in the context of FO theories.
Important among these are syntactic characterizations of FO theories
in $PSC(k)$ and in $PSC_f$, and an understanding of whether $PSC_f$
strictly subsumes $PSC$ for FO theories.  It is also interesting that
the semantic notions of $PSC(k)$ and $PCE(k)$ remain non-trivial over
classes of finite structures.  This contrasts with other semantic
notions (like those mentioned above) that have been used earlier to
characterize $\Sigma^0_2$ and $\Pi^0_2$ over arbitrary structures, but
reduce to trivial properties over any class of finite structures (see
Appendix \ref{appendix:comparison} for details).  This motivates
investigating classes of finite structures over which $PSC(k)$ and
$PCE(k)$ semantically characterize $\exists^k\forall^*$ and
$\forall^k\exists^*$ sentences respectively.  Some such classes were
considered in ~\cite{wollic12-paper}.  Further investigations in this
direction would be a natural extension of recent work on preservation
theorems over special classes of finite
structures~\cite{dawar-pres-under-ext,dawar-hom}.\\

\vspace{-3pt}
{\fontsize{11}{13} \selectfont {\tbf{Acknowledgements}}}: We are very
grateful to Anuj Dawar, Rohit Parikh and Anand Pillay for valuable
discussions and feedback.

\vspace{-10pt}

\bibliography{refs} \bibliographystyle{splncs}
\appendix

\section{Comparing  the notions of $PSC$ and $PCE$ with other related notions in the literature}\label{appendix:comparison}

We present three notions from the literature that provide
characterizations of $\Sigma^0_2$ and $\Pi^0_2$ over arbitrary
structures. None of these characterizations relate the count of
quantifiers in the $\Sigma^0_2$ and $\Pi^0_2$ sentences to any
quantitative property of their models. As a consequence, none of these
yield the {\lt} theorem as a special case. We also show below that all
the three notions become trivial over any class of finite
structures. \\

1. \emph{Preservation under unions of ascending chains}:

An \emph{ascending chain} is a sequence of structures $M_1, M_2,
\ldots$ such that $M_1 \subseteq M_2 \subseteq \ldots$. Given an
ascending chain $C$, the union of the chain is the unique structure
$N$ such that (a) the union of the universes of the structures in $C$
is exactly the universe of $N$ (b) $M \subseteq N$ for each $M \in C$.
A sentence $\phi$ is said to be \emph{preserved under unions of
  ascending chains}, if for every chain $C$, if all structures in $C$
model $\phi$, then the union of $C$ also models $\phi$. A classical
theorem states that over arbitrary structures, $\phi$ is preserved
under unions of ascending chains iff $\phi$ is equivalent to a
$\Pi^0_2$ sentence~\cite{chang-keisler}.  

Now consider any class $\mc{P}$ of finite structures and consider an
ascending chain of structures from $\mc{P}$. One can check that either
the union of the chain is not a finite structure or it is the same as
some structure in the chain. Then over $\mc{P}$, any sentence is
preserved under unions of ascending chains!\\

2. \emph{Preservation under intersections of descending chains}:

An \emph{descending chain} is a sequence of structures $M_1, M_2,
\ldots$ such that $M_1 \supseteq M_2 \supseteq \ldots$. Given an
descending chain $C$, the intersection of the chain is a structure $N$
such that (a) the intersection of the universes of the structures in
$C$ is exactly the universe of $N$ (b) $N \subseteq M$ for each $M
\in C$.  Note that $N$  exists iff the intersection of the
universes of the structures in $C$ is non-empty. Further, it is unique
if it exists.  A sentence $\phi$ is said to be \emph{preserved under
intersections of descending chains}, if for every chain $C$, if all
structures in $C$ model $\phi$, then the intersection of $C$, if it
exists, also models $\phi$. A classical theorem states that over
arbitrary structures, $\phi$ is preserved under intersections of
descending chains iff $\phi$ is equivalent to a $\Pi^0_2$
sentence~\cite{chang-keisler}.  

Now consider any class $\mc{P}$ of finite structures and consider a
descending chain of structures from $\mc{P}$. One can check that the
intersection of the chain must necessarily be the same as some
structure in the chain. Then over $\mc{P}$, any sentence is preserved
under intersections of descending chains!\\

3. \emph{Preservation under 1-sandwiches}:

A notation before proceeding: By $M \preceq N$, we mean that $M$ is an
elementary substructure of $N$.  

Given structures $M$ and $N$, we say $M$ 1-sandwiches $N$ if there
exist structures $M'$ and $N'$ such that (i) $M \preceq M'$ (ii) $N
\preceq N'$ and (iii) $M \subseteq N' \subseteq M'$. We say that
$\phi$ is preserved under 1-sandwiches if it is the case that if $N
\models \phi$ and $M$ 1-sandwiches $N$, then $M \models \phi$. A
classical theorem states that $\phi$ is preserved under 1-sandwiches
iff it is equivalent to a $\Pi^0_2$ sentence~\cite{chang-keisler}.

Since one can capture a finite structure upto isomorphism using a
single FO sentence, it follows that given two finite structures $M$
and $M'$, $M \preceq M'$ implies that $M$ is isomorphic to $M'$. Now
consider any class $\mc{P}$ of finite structures and let $M$ and $N$
be structures from $\mc{P}$. Then if $M$ 1-sandwiches $N$, it follows
from our observation above that $M$ must be isomorphic to $N$. Once
again then, any sentence is preserved under 1-sandwiches!\\

In contrast to the three notions considered above, consider the
notions of $PSC(k)$ and $PCE(k)$ introduced in~\cite{wollic12-paper}
and in the present paper. Because these notions allow us to obtain
preservation theorems (Theorems \ref{theorem:PSC(k)=E^kA*}
and \ref{theorem:PCE(k)=A^kE*}) that relate the count of quantifiers
in the leading block of quantifiers of $\exists^* \forall^*$ and
$\forall^* \exists^*$ sentences to quantitative properties of their
models, we obtain the substructual and extensional versions of the
{\lt} theorem as special cases. In addition, the notions of $PSC(k)$
and $PCE(k)$ remain non-trivial over the class of all finite
structures, amongst other classes of finite structures. In other
words, for each $k$, there are atleast two sentences such that, over
the class of all finite structures, one of these sentences is in
$PSC(k)$ and the other is not (likewise for $PCE(k)$). Any sentence in
$PSC(k)$ over arbitrary structures would also be in $PSC(k)$ over all
finite structures (likewise for $PCE(k)$). As an example of a sentence
that is not in $PSC(k)$, consider the sentence $\phi = \forall
x \exists y E(x, y)$.  Let $G$ be a cycle of length $k+1$. This is a
model of $\phi$. However, no proper induced subgraph $G'$ of $G$ is a
model of $\phi$ since $G'$ must contain a vertex which has no outgoing
edge. Then the only core of $G$ is the set of all vertices of $G$ --
but this has size $k+1$. This shows that $\phi \notin PSC(k)$. (In
fact, this argument shows that $\phi \notin PSC$).  Likewise, by
Corollary \ref{corollary:PSC(k)-PCE(k)-duality-for-sentences}, $\psi =
\exists x \forall y E(x, y) \notin PCE(k)$ (in fact, $\psi \notin PCE$).
This motivates studying classes of finite structures over which
$PSC(k)$ and $PCE(k)$ do form semantic characterizations of the
$\Sigma^0_2$ and $\Pi^0_2$ classes of FO (the notions in the
literature mentioned above surely cannot give us characterizations for
these classes). Indeed, characterizations of $\Sigma^0_2$ using
$PSC(k)$ were obtained over some interesting classes of finite
structures in~\cite{wollic12-paper}.

\end{document}